\documentclass[a4paper,11pt]{article}
\pdfoutput=1
\usepackage[dvipsnames]{xcolor}
\usepackage[T1]{fontenc} 
\usepackage[toc,page]{appendix}
\usepackage{diagbox}
\usepackage{comment}
\usepackage{jheppub}
\usepackage{circledsteps}
\usepackage{subcaption}
\usepackage{natbib}
\setcitestyle{square,comma,numbers,sort&compress}
\usepackage{tabularx}
\usepackage{enumerate}
\usepackage{mathrsfs}
\usepackage{tabularx,booktabs}
\usepackage{caption}
\usepackage[compat=1.1.0]{tikz-feynman} 
\usepackage{cancel}
\usepackage{slashed}

\usepackage[section]{placeins}
\usepackage{cleveref}

\usepackage{simpler-wick}
\newcolumntype{Y}{>{\centering\arraybackslash}X}

\newcommand{\be}{\begin{equation}}
\newcommand{\ee}{\end{equation}}
\newcommand{\bea}{\begin{eqnarray}}
\newcommand{\eea}{\end{eqnarray}}

\definecolor{darkgreen}{rgb}{0,0.5,0}

\definecolor{myRED}{rgb}{0.8, 0.25, 0.33}

\begin{document}

\title{\huge Gauged Global Strings}

\author{Xuce Niu,}
\author{Wei Xue,}
\author{and Fengwei Yang}

\affiliation{Institute for Fundamental Theory, Department of Physics,
University of Florida,\\Gainesville, FL 32611, USA }
\emailAdd{xuce.niu@ufl.edu}
\emailAdd{weixue@ufl.edu}
\emailAdd{fengwei.yang@ufl.edu}

\abstract{
We investigate the string solutions and cosmological implications of the gauge ${\rm U(1)_Z}\,\times$ global ${\rm U(1)_{PQ}}$ model.
With two hierarchical symmetry-breaking scales, the model exhibits three distinct string solutions: 
a conventional global string, a global string with a heavy core, and a gauge string 
as a bound state of the two global strings.
This model reveals rich phenomenological implications in cosmology. During the evolution of the universe, 
these three types of strings can form a Y-junction configuration. 
Intriguingly, when incorporating this model with 
the QCD axion framework, 
the heavy-core global strings emit more axion particles compared to conventional axion cosmic strings due to their higher tension.
This radiation significantly enhances the QCD axion dark matter abundance, thereby opening up
the QCD axion mass window. Consequently, 
axions with masses exceeding $\sim 10^{-5}\,  {\rm eV}$ have the potential to constitute the whole dark matter abundance.
Furthermore, in contrast to 
conventional gauge strings, the gauge strings in this model exhibit a distinctive behavior by radiating axions.
}
\maketitle
\flushbottom

\section{Introduction}

Cosmic strings, soliton solutions in field theory, arise when loops in the vacuum manifold cannot be contracted to a point 
\cite{Kibble:1976sj,Vilenkin:1984ib}.
In the context of symmetry-breaking, where a symmetry group $G$ is spontaneously broken to a subgroup $H$, $G \to H$, the vacuum manifold 
is a quotient space, $G/H$.
Cosmic string solutions are connected to the non-trivial first homotopy group $\pi_1 ( G/ H)$.
The simplest cosmic string originates from a complex scalar field, denoted as $\Phi$, with a global $\rm U(1)$ symmetry. 
Its vacuum expectation value~(VEV), $\langle \Phi \rangle = \frac{1}{\sqrt{2}} f_a$ spontaneously breaks its global 
$\rm U(1)$ symmetry. A cosmic string along the $z$-direction exhibits a scalar field configuration in polar coordinates $(r, \theta )$,
\begin{equation}
   \Phi (r, \theta ) = \frac{1}{\sqrt{2}} f_a  {\rm e}^{ i\theta} \, , \quad  r \to \infty
\end{equation}
with a winding number $1$. 
The energy per unit length of the string is estimated by integrating radially from the inverse of the scalar mass 
$m^{-1}$ to a long distance cutoff $L$, 
\begin{equation}
   \mu \simeq   2 \pi \int_{m^{-1}}^L  {\rm d}r \frac{1}{r}  |  {\partial_\theta } \Phi (r, \theta ) |^2 
       = \pi f_a^2 \ln ( m L) 
   \label{eq:mu0}
\end{equation}
The tension of a global string is predominantly contributed by the gradient term outside of the string core. 
Alternatively, considering an Abelian Higgs model where the $\rm U(1)$ symmetry is a gauge symmetry,
the cosmic strings, known as gauge strings, exhibit finite tension concentrated inside the string core. 
While the gauge strings can have similar scalar configurations as global strings, the energy from the gradient term 
outside the core regime is eliminated by a gauge configuration.

Cosmic strings generically form from a phase transition in the early universe through the Kibble mechanism \cite{Kibble:1976sj}.
The broken symmetry is restored at the high temperature of the universe, $T \gtrsim f_a$. A phase 
transition occurs when the temperature falls around $f_a$. During this transition, the vacuum expectation value $\langle \Phi \rangle$
turns on, and the symmetry is spontaneously broken. The phase of $\langle \Phi \rangle$ in the vacuum manifold is chosen at random
beyond the correlation length of the phase transition, resulting in the formation of cosmic strings. 
Subsequently, the interactions between cosmic strings 
lead to a few long strings per Hubble volume, entering a string scaling regime.

Cosmic string networks in the universe provide intriguing signatures, and their detection is an exciting direction
to probe UV physics. 
The existence of cosmic strings influences the large-scale structure of 
the universe \cite{Zeldovich:1980gh,Vilenkin:1981iu,Kibble:1980mv,Pen:1997ae,Magueijo:1995xj}. The current constraint on the cosmic string tension arises from analyzing the angular power spectrum of cosmic microwave
background~(CMB) \cite{Planck:2013mgr,Urrestilla:2011gr,Lizarraga:2014xza,Lazanu:2014xxa}. Furthermore, a cosmic string loop cannot survive in the universe forever. Gauge strings emit gravitational radiations
from loop oscillations, detectable by current and future gravitational wave detectors \cite{LIGOScientific:2021nrg,Ellis:2020ena,Blanco-Pillado:2021ygr,Auclair:2019wcv,Boileau:2021gbr,Punturo:2010zz,Yagi:2011wg,AEDGE:2019nxb,Hild:2010id,Sesana:2019vho,2017arXiv170701348T}. Additionally, considering interactions between cosmic string and particles, such as photons, 
opens new possibilities for detection in the CMB or astrophysical observations\cite{Agrawal:2019lkr,Agrawal:2020euj, Jain:2021shf, 
Yin:2021kmx,Jain:2022jrp, Yin:2023vit,Hook:2023smg,Hagimoto:2023tqm}.

One well-motivated cosmic string is the QCD axion string, as a global string. 
The QCD axions \cite{Weinberg:1977ma,Wilczek:1977pj,Shifman:1979if,Kim:1979if,Zhitnitsky:1980tq,Dine:1981rt, Preskill:1982cy,Abbott:1982af,Dine:1982ah}
provide a physically intriguing solution, solving the strong CP problem  \cite{Peccei:1977hh,Peccei:1977ur}, and 
serving as a dark matter candidate \cite{Preskill:1982cy,Abbott:1982af,Dine:1982ah}.
Axion string emission to axions can be a dominant contributor to dark matter 
abundance, though the emitted axion energy spectrum is still an unsettled question \cite{Davis:1989nj,Dabholkar:1989ju,Hagmann:1990mj,Battye:1993jv,Battye:1995hw,Chang:1998tb,Yamaguchi:1998iv,Yamaguchi:1998gx,Yamaguchi:1999yp,Yamaguchi:1999dy,Hagmann:2000ja,Martins:2003vd,Wantz:2009it,Hiramatsu:2010yn,Hiramatsu:2010yu,Hiramatsu:2012gg,Hiramatsu:2012sc,Kawasaki:2014sqa}. Recently, numerical simulations \cite{Klaer:2017ond,Gorghetto:2018myk,
Buschmann:2019icd,Gorghetto:2020qws,Buschmann:2021sdq} 
make efforts to address axion dark matter abundance from the QCD axion string radiation, 
predicting an axion mass of $\mathcal{O}(10-100)\, \mu {\rm eV}$ 
to explain the full dark matter abundance. 
The QCD axions motivate world-wide efforts for their search, 
not only focusing on a specific mass range but also planning to cover a 
broad mass range in the future experiments \cite{Stern:2016bbw,Alesini:2020vny,DeMiguel:2023nmz,Lawson:2019brd,Beurthey:2020yuq,McAllister:2017lkb,Aja:2022csb,brass,BREAD:2021tpx}. 
Also, various mechanisms,
including parametric resonance \cite{Agrawal:2017eqm,PhysRevLett.120.211602,Harigaya:2019qnl}, 
anharmonicity effect \cite{PhysRevD.33.889,PhysRevD.45.3394,PhysRevD.80.035024},
domain walls decays \cite{PhysRevLett.48.1156,PhysRevD.59.023505,Takashi_Hiramatsu_2011,Hiramatsu:2012sc,PhysRevD.91.065014,PhysRevD.94.049908,HARIGAYA20181,PhysRevD.100.063530},
axion production during a kination era \cite{PhysRevD.81.063508}, 
and kinetic misalignment \cite{Co:2019jts},
allow for different axion masses to explain the dark matter abundance. 
Here, we provide another mechanism from cosmic string decays.

In this paper, we introduce a simple model having two complex fields, $\Phi_1$ and $\Phi_2$,  with a symmetry of 
gauge $\rm U(1)_Z \,\times$  global $\rm U(1)_{PQ}$. 
From this model, we obtain two kinds of global string solutions and one gauge string solution from the spontaneous symmetry breaking,
as the three most energetically favorable string configurations. 
The two global strings arise from 
the winding around $\Phi_1$ and $\Phi_2$, respectively. Assuming that the vacuum expectation values of the two fields are hierarchical, 
we observe that one global string is heavier than another. The lighter global string tension is close to the conventional
QCD axion string. 
The third string is a bound gauge string formed by combining the two global strings together. 
With the hybrid string solutions, we investigate their cosmological implications.
The system displays unique cosmological dynamics and signatures. This is evident in the formation and evolution of string networks, as well as in the radiation emitted by string loops.
In the context of QCD axion physics, heavier global strings emit more axions due to their larger tension, contributing more 
to the axion dark matter abundance.
The large tension has been used to emulate the behavior of axion strings in a cosmological simulation \cite{Klaer:2017qhr}.
Additionally, for a gauge string from the Abelian Higgs model, gravitational
radiation is the dominant channel for the comic strings to lose energy. However, the gauge string in this model couples to massless Goldstone modes,
raising the intriguing question of which particles the gauge strings radiate dominantly.

The structure of this paper is organized as follows. \Cref{sec:2} introduces the ${\rm U(1)_Z}$ $\times\, {\rm U(1)_{PQ}}$ model, and then we embed the model into the QCD axion framework. Following this, the section provides the string solutions for this model, which are further validated through numerical analysis in \cref{sec:3}. 
The cosmological implications of the model are explored in \cref{sec:4}, and our findings are summarized in \cref{sec:conclusion}.

\section{${\rm U}(1)_{\rm Z} \times {\rm U}(1)_{\rm PQ}$}
\label{sec:2}

In this section, we introduce a model having a gauge symmetry ${\rm U}(1)_{\rm Z}$ and a global symmetry ${\rm U}(1)_{\rm PQ}$.
The dynamics of this model are driven by two scalar fields, $\Phi_1$ and $\Phi_2$, which will break the symmetry sequentially 
at distinct energy scales in the cosmic evolution. 
Furthermore, we consider the intriguing possibility of integrating this model with the QCD axion framework, 
so that the full model merges  
the Abelian gauge symmetry with the original QCD axion models \cite{Kim:1979if,Shifman:1979if,Dine:1981rt,Zhitnitsky:1980tq,DiLuzio:2020wdo}.

\subsection{The model}
\label{sec:model}

We consider the gauge symmetry and the global symmetry, ${\rm U}(1)_{\rm Z} \times {\rm U}(1)_{\rm PQ}$, both of which are 
broken by two complex scalar $\Phi_1$ and $\Phi_2$.
The dynamics are described by their gauge-invariant and renormalizable Lagrangian density, which takes the form
\begin{equation}
   {\cal L} = -\frac{1}{4} Z_{\mu \nu} Z^{\mu \nu}  +
      D_\mu \Phi_1^\dagger D^\mu \Phi_1 - \frac{\lambda_1}{4} \left( |\Phi_1|^2 - \frac{v_1^2}{2} \right)^2 
      +
     D_\mu \Phi_2^\dagger D^\mu \Phi_2 - \frac{\lambda_2}{4} \left( |\Phi_2|^2 - \frac{v_2^2}{2} \right)^2  \, .
\label{eq:Lag}
\end{equation}
$Z^{\mu \nu}$ represents the field strength of the $\rm U(1)_Z$ gauge boson $Z_{\mu}$,  
defined as $Z^{\mu \nu} = \partial^\mu Z^{\nu} - \partial^\nu Z^\mu$.
The scalar fields interact with the gauge boson $Z_{\mu}$ with the gauge coupling $e$ 
through the covariant derivative terms, as expressed by
\begin{equation}
   D_\mu \Phi_i=(\partial_\mu-i e Z_\mu)\Phi_i,~~i=1,2 \, . 
\end{equation}
We assign that $\Phi_1$ carries charges $(+1, +1) $ under ${\rm U}(1)_{\rm Z} \times {\rm U}(1)_{\rm PQ}$, 
while $\Phi_2$ has charges $(+1, -1) $. 
It is important to note that the charges cannot be entirely determined by the above Lagrangian density. They should be determined by 
the scalars' coupling to other particles or a UV theory.
The charges of $\Phi_1$ and $\Phi_2$ can be extended to other values, 
and the implication for cosmic strings will be discussed in \cref{sec:str-tension,sec:formation-string-network}. 
For simplicity, we consider two independent Mexican-hat potentials for the scalar fields with self-couplings $\lambda_1$ and $\lambda_2$ 
while assuming that interactions between the two scalars, namely $|\Phi_1|^2 |\Phi_2|^2$,  are either negligible or zero.
In the vacuum, $\Phi_1$ and $\Phi_2$ acquire non-zero expectation values, denoted as $v_1$ and $v_2$,
\begin{equation}
   \langle\Phi_1\rangle=\frac{v_1}{\sqrt{2}},
      ~\langle\Phi_2\rangle=\frac{v_2}{\sqrt{2}} \, . 
\end{equation}
This spontaneous symmetry breaking leads to the sequential breaking of the $\rm U(1)_Z \times U(1)_{PQ}$ symmetry 
in the cosmic evolution, particularly when $v_1 \gg v_2$.
The first phase transition results in a non-zero VEV of $ \langle \Phi_1 \rangle $, leading to the breaking of $\rm U(1)_Z$ 
and giving the gauge boson $Z^\mu$ mass, $m_Z = e v_1 $.
After the second phase transition, the non-zero VEV of $ \langle \Phi_1 \rangle $ and $ \langle \Phi_2 \rangle $ further breaks the 
global symmetry $\rm U(1)_{PQ}$, with the gauge boson mass increasing, $m_Z = e \sqrt{v_1^ 2+ v_2^2}$.

We conveniently parametrize the perturbations of $\Phi_1$ and $\Phi_2$ using real scalar fields 
$ \tilde\phi_1 (x)$, $\tilde\phi_2(x)$, $\pi_1(x)$ and $\pi_2 (x)$,
\begin{equation}
   \begin{split}
\Phi_1 (x) &= \frac{1}{\sqrt{2}}  \phi_1 (x)  \, e^{i \alpha_1 }
 = \frac{1}{\sqrt{2}} ( v_1 + \tilde\phi_1 (x) ) \, e^{i \pi_1 (x) / v_1 }  
   \, ,
   \\
 \Phi_2 (x) &= \frac{1}{\sqrt{2}}  \phi_2 (x)  \, e^{i \alpha_2 }
 = \frac{1}{\sqrt{2}} ( v_2 + \tilde\phi_2 (x) ) \, e^{i \pi_2 (x) / v_2 } \, ,
   \end{split} 
   \label{eq:phi12para}
\end{equation}
where $\alpha_1$ and $\alpha_2$ represent the rotation angle of $\Phi_1$ and $\Phi_2$, respectively.
We identify the axion, denoted as $a(x)$, by ensuring that it is orthogonal to the would-be Goldstone boson $\pi_z$ 
or the longitudinal mode of the gauge boson $Z^\mu$. Using the expression of the $\rm U(1)_Z$ current $J_z^\mu$ and 
$\pi_z$-to-vacuum matrix element, 
$  \langle {\rm vac} | J_z^{\mu} (0) | \pi_z (p) \rangle = i p^\mu v $, 
we identify the would-be Goldstone boson of $\rm U(1)_Z$, which takes the form  
\begin{equation}
   \pi_z (x) = \frac{1}{v} \left( v_1 \, \pi_1 + v_2 \, \pi_2 \right) \, , 
\end{equation}
where 
$v= \sqrt{v_1^ 2+ v_2^2} $.  
Requiring orthogonality between the axion field and $\pi_z$ yields the expression for $a(x)$
\begin{equation}
   a (x) = \frac{1}{v} \left( v_2 \, \pi_1 - v_1 \, \pi_2 \right) \, .
\label{eq:api12}
\end{equation}
According to the symmetry of $\Phi_{1}$ and $\Phi_2$, we can express $\pi_1$ and $\pi_2$
in terms of the rotation angles of $\rm U(1)_{Z}$ and $\rm U(1)_{PQ}$ in the vacuum manifold, 
specifically 
\begin{equation}
\pi_1 / v_1 = \alpha_Z + \alpha_{\rm PQ} \, , \quad \pi_2 / v_2 = \alpha_Z - \alpha_{\rm PQ} 
\, . 
\end{equation}
Here, $\alpha_Z$ and $\alpha_{\rm PQ}$ represent the rotation angles of $\rm U(1)_{Z}$ and $\rm U(1)_{PQ}$, respectively. 
Consequently, we can rewrite the axion fields in terms of the $\rm U(1)_{PQ}$ rotation angle $\alpha_{\rm PQ} $,
\begin{equation}
   a(x) = v_a \alpha_{\rm PQ} \, ,  \quad   v_a = \frac{2 v_1 v_2 } {\sqrt{v_1^2 + v_2^2} } \, .
\end{equation}
The parameter $v_a$ tells us the magnitude of the vacuum expectation value that spontaneously breaks $\rm U(1)_{PQ}$.

\subsection{QCD axion}
\label{sec:QCDaxion}

The ${\rm U}(1)_{\rm Z} \times {\rm U}(1)_{\rm PQ}$ model provides an elegant framework for realization, 
and it becomes particularly intriguing when integrated into QCD axion models. 

One possible approach, based on the KSVZ model, involves introducing a vector-like heavy fermion
that can be decomposed into its left-handed and right-handed components, ${\cal Q} = {\cal Q}_L + {\cal Q}_R$. 
This fermion resides in the fundamental representation of the Standard Model $\rm SU(3)_c$ color symmetry and is a singlet 
under the $\rm U(1)_Z$ symmetry and other Standard Model symmetries.
Moreover, the fermion carries a chiral charge under the PQ symmetry, with ${\cal Q}_L$ having a $+1$ charge 
and ${\cal Q}_R$ a $-1$ charge.
Since it is a singlet in the gauge $\rm U(1)_Z$ symmetry, the fermion does not introduce any additional $\rm U(1)_Z$ anomaly, 
preserving the gauge symmetry anomaly-free.
 
\begin{figure}[!ht]
    \centering
     \includegraphics[width=0.8\textwidth]{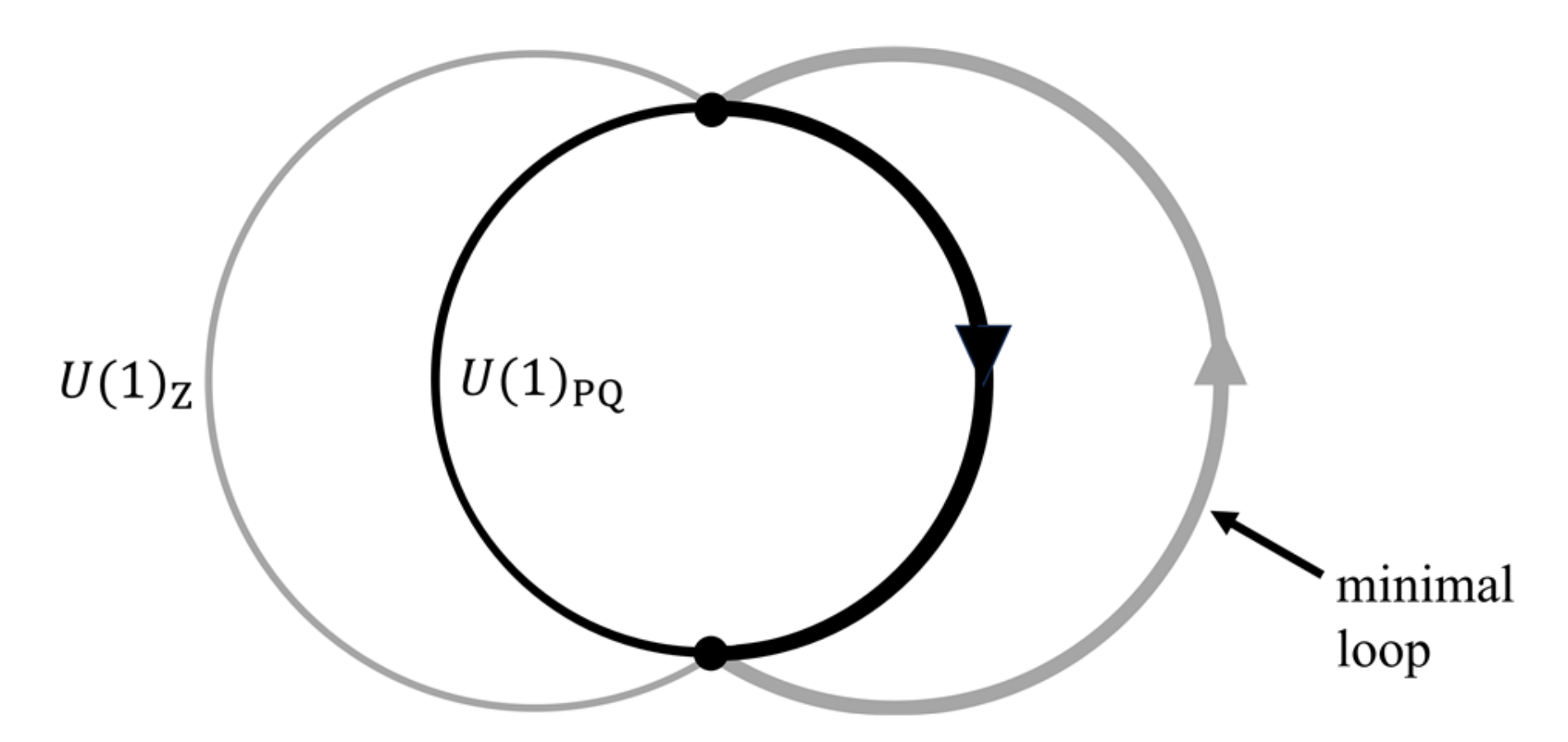}
     \caption{\label{fig:vacuum-manifold}The cross-section of the vacuum manifold of ${\rm U(1)_Z\times U(1)_{PQ}}$, where the two black dots represent the angle $\alpha_{\rm PQ}=0$ and $\alpha_{\rm PQ}=\pi$.}
\end{figure}

Taking into account the symmetry of the fermion $\cal Q$, we construct the Lagrangian
\begin{equation} 
   {\cal L} = i \bar{ \cal Q}  \slashed{D} {\cal Q}  - \frac{y}{\Lambda} \left(  \Phi_1 \Phi_2^* \bar{ \cal Q}_L {\cal Q}_R  + h.c.\right)  \, ,
\end{equation} 
where $\Lambda$ represents the UV cutoff of the theory. The expectation values of $\Phi_1$ and $\Phi_2$ yield the fermion mass,
$m_{\cal Q} = y v_1 v_2 / ( 2 \Lambda )$. 
By integrating out the heavy scalars and employing \cref{eq:phi12para,eq:api12}, 
we derive the 
effective Lagrangian for axion couplings to fermions
\begin{equation}
   {\cal L} \supset - m_{\cal Q} e^{ i 2 a /v_a } \bar{\cal Q}_L {\cal Q}_R + h.c. \, .
\end{equation}
Subsequently, we can deduce axion interactions with the Standard Model particles, 
including axion couplings to gluons. The derivation and results parallel those of the KSVZ model.
Through a field-dependent axial transformation, 
\begin{equation}
   \mathcal{Q}_L\rightarrow e^{i a / v_a }\mathcal{Q}_L\, , \quad 
   \mathcal{Q}_R\rightarrow e^{-i a / v_a }\mathcal{Q}_R,
\end{equation}
the heavy fermion becomes disentangled from the axion, introducing an axion-gluon coupling term
\begin{equation}
   {\cal L} \supset 
      \frac{g_s^2}{16\pi^2}
      \frac{ a} {v_a  } 
      G^a_{\mu\nu}\tilde{G}^{a}_{\mu\nu} \,  
      = 
         \frac{g_s^2}{32\pi^2}
    \frac{a} {f_a } 
      G^a_{\mu\nu}\tilde{G}^{a}_{\mu\nu} \,  , 
\label{eq:agg}
\end{equation}
where $g_s$ is the coupling constant of $\rm SU(3)_c$. The axion-gluon coupling defines 
the axion decay constant $f_a$. Notably, in this model, $v_a = 2 f_a$. 
The ratio of $v_a$ and $f_a$, often referred to as the domain wall number, is represented by 
$ \frac{v_a}{f_a}$.
However, in this case, the domain wall number for cosmic string solutions is $N = 1$.
This arises because, in the vacuum manifold of $\rm  U(1)_{PQ}$, the minimal winding is achieved by choosing the angle $\alpha_{\rm PQ}$ 
from $0$ to $\pi$. $\alpha_{\rm PQ} = 0 $ and $\alpha_{\rm PQ} = \pi$ are gauge-equivalent, and $0$ can return to $\pi$ through the 
gauge group $ \rm U(1)_Z$ (see \cref{fig:vacuum-manifold}). A similar method of counting domain wall number is observed in the PQWW model 
\cite{Peccei:1977hh,Weinberg:1977ma,Wilczek:1977pj}, which has three domain walls in an axion string, but $\frac{v_a}{f_a}  = 6$.

Alternatively, we can construct a model by introducing two sets of quarks, denoted as ${\cal Q}_{1}$ and ${\cal Q}_{2}$, 
as discussed by Barr and Seckel \cite{Barr:1992qq}. 
These quarks are color-triplets under $\rm SU(3)_c$ and are assigned the following chiral charges under $\rm U(1)_Z \times U(1)_{PQ}$:
${\cal Q}_{1L}$ has charges $(  1/2, 1/2)$,
${\cal Q}_{1R}$ has charges $(  - 1/2, -1/2)$,
${\cal Q}_{2L}$ has charges $(  - 1/2,  - 1/2)$, and 
${\cal Q}_{2R}$ has charges $(   1/2, 1/2)$.
This charge assignment ensures that $\rm U(1)_Z$ remains anomaly-free.
Furthermore, 
it leads to ${\cal Q}_1$ interacting with $\Phi_1$ and ${\cal Q}_2$ interacting with $\Phi_2$ though the Lagrangian density, 
\begin{equation}
   {\cal L} =  \Phi_1  \bar{ \cal Q}_{1L} {\cal Q}_{1R} 
       + \Phi_2  \bar{ \cal Q}_{2L} {\cal Q}_{2 R}  + h.c. \, .
\end{equation}
Following a similar procedure, we arrive at the same axion gluon coupling shown in \cref{eq:agg}.

Let us comment on other extensions of these models.
In both realizations, we can introduce $N_f$ flavors of the heavy quarks, thereby suppressing the axion decay constant $f_a$ as a factor of $1/N_f$. 
This increases domain wall numbers in the cosmic string solutions. Additionally, in the Standard Model, 
$\rm U(1)_{B-L}$ is anomaly-free and could serve as a gauge symmetry. Another intriguing possibility is that $\rm U(1)_Z$ symmetry can be 
identified with the $\rm U(1)_{B-L}$ \cite{Ibe:2018hir}.

\subsection{String solutions\label{sec:str-tension}}

In this section, we study the string solutions of ${\rm U(1)_Z}\times {\rm U(1)_{PQ}}$ analytically. 
Initially, we analyze the string solutions beyond their core regions to pinpoint the three most energetically favorable string configurations.
Then, we extend our analysis to include string solutions with arbitrary winding numbers $(j,k)$.
Finally, we extrapolate our findings to encompass generic gauge charges.

\subsubsection*{The gradient energy of $(1,0)$, $(0,1)$ and $(1,1)$ strings}

Before delving into the total energy per unit length of strings, we focus on the gradient term for $(1,0)$, $(0,1)$ and $(1,1)$ strings.  
Examining the gradient energy allows us to distinguish the $(1,0)$, $(0,1)$ strings are global strings, while 
$(1,1)$ strings are gauge strings. Also, these three string configurations are the lightest ones, playing a major role in cosmology.

For the winding number $(1,0)$, the classical configurations of the scalar and gauge fields outside the string cores take the form 
\begin{equation}
  \Phi_1  = \frac{1}{\sqrt{2} } v_1 \, e^{i \theta }  \, , \quad 
  \Phi_2  = \frac{1}{\sqrt{2} } v_2 \, , \quad 
   Z_\mu = c \, \partial_\mu \theta \, , \quad  r \to \infty
\end{equation}
Here, $c$ is a normalization of the gauge field, determined by minimizing the gradient energy of strings.
The gradient energy per unit length is then calculated by integrating the 2D cross-section of the string, yielding
\begin{equation}
   \begin{split}
   \mu_{k, (1,0)} &=  \int_0^{2 \pi} {\rm d } \theta \int_\delta^L {\rm d} r   \, r
      \left( | ( \frac{ 1}{r} \partial_\theta  - i e Z_\theta  ) \Phi_1|^2
         +
      |   (- i e Z_\theta  ) \Phi_2|^2
      \right)
      \\
   &=  \pi \ln (\frac{L }{\delta} ) \left[ v_1^2 ( 1 - e c )^2  + v_2^2 (ec)^2 \right]
   \end{split}
\end{equation}
where we take the core size as $\delta$. Since $m_Z$ is typically lighter than the scalar field mass $m_1$ and $m_2$, 
we can choose the core size $\delta \simeq m_Z^{-1}$.\footnote{
We consider the gradient term outside the core region. There are also gradient corrections from the radius of $m_1^{-1}$ to $m_Z^{-1}$,
being neglected here but included in the full tension calculation.
The contribution is about $\pi v_1^2\ln(m_1/m_Z)$. It can be dominant when $v_1\gg v_2$.} 
The parameter $c$ is determined by minimizing the tension,
\begin{equation}
   c = \frac{1}{e} \frac{ v_1^2} { v_1^2+ v_2^2} \, .  
\end{equation}
Therefore, the $(1,0)$ tension outside the core is 
\begin{equation}
  \mu_{k, (1,0)} = \pi \frac{ v_1^2 v_2^2} {v_1^2+ v_2^2} \ln (\frac{L }{\delta} )  
   = \pi f_a^2 \ln (\frac{L }{\delta} )   
\end{equation}

Next, for the $(0,1)$ string tension outside the string core, the configurations take the form 
$  \Phi_1  = \frac{1}{\sqrt{2} } v_1$, $\Phi_2  = \frac{1}{\sqrt{2} } v_2  e^{i \theta } $ and $ Z_\mu = c \, \partial_\mu \theta $.
$c= \frac{1}{e} \frac{ v_2^2} { v_1^2+ v_2^2}$ minimizes the energy of the $(0,1)$ string. The result should be the same as $(1,0)$ string
by switching $v_1$ and $v_2$. 
The tension of $(0,1)$ string is the same as the $(1,0)$ string,
\begin{equation}
  \mu_{k, (0,1)} 
   = \pi f_a^2 \ln(\frac{L }{\delta} )   
\end{equation}
There is an intuitive way to understand that the $(1,0)$ and $(0,1)$ strings share the same tension outside the core regimes. 
The $(1,0)$ string configuration outside core is equivalent 
to a $(0,-1)$ string through a gauge transformation, $\alpha_Z \to \alpha_Z+ \theta$. Consequently, 
$(1,0)$ and $(0,1)$ are string and anti-string.\footnote{We thank Pierre Sikivie for the discussion on this point.}

The $(1,1)$ string results from combing a $(1,0)$ string with its anti-string, $(0,1)$ string, implying 
zero tension outside the core.
This result aligns with our study of field configurations.
Using the same energy minimization procedure, we find that the $Z^\mu $ configuration with $c=1/e$ cancels
the $\Phi_1$ and $\Phi_2$ gradient terms simultaneously. Hence, we conclude that
the tension outside the core of $(1,1)$ string is zero, $\mu_{k, (1,1)}  =0 $. Consequently, the $(1,1)$ string is identified as a gauge string.

\subsubsection*{The tension of $(j,k)$ strings}
By considering the time-independent stable field configuration and choosing the gauge $Z_0=0$, 
the full string tension is obtained by the two-dimensional spatial integral of the Hamiltonian density, 
\be
\label{tension}
\mu=\int d^2x \left[\frac{B_z^2}{2}+|D_i\Phi_1|^2+|D_i\Phi_2|^2+\frac{\lambda_1}{4}\left(|\Phi_1|^2-\frac{v_1^2}{2}\right)^2+\frac{\lambda_2}{4}\left(|\Phi_2|^2-\frac{v_2^2}{2}\right)^2\right],
\ee
where $B_z=\frac{1}{r}\frac{d(rZ_\theta(r))}{dr}$ is the $z$-component of the ``magnetic'' field  ${\bf B}=\nabla\times {\bf Z}$.

For a $(j,k)$ string, we choose the {\it Ansatz} for the fields: 
\bea
\Phi_1(r,\theta)&=&\frac{v_1}{\sqrt{2}}f_1(r)e^{ij\theta},\nonumber\\
\Phi_2(r,\theta)&=&\frac{v_2}{\sqrt{2}}f_2(r)e^{ik\theta},\nonumber\\
Z_\theta(r)&=&c\frac{g(r)}{r},\label{eq:Z-ansatz}
\eea
where $f_1(r),~f_2(r)$, and $g(r)$ are the profile functions and where we have $c\equiv \frac{1}{e}\frac{jv_1^2+kv_2^2}{v_1^2+v_2^2}$ to 
minimize the gradient energy (deviation in appendix \ref{App:A}). They need to satisfy the boundary conditions when $j\neq0$ and $k\neq0$:
\bea
&~&f_{1,2}(r\rightarrow 0)\rightarrow 0,~f_{1,2}(r\rightarrow\infty)\rightarrow 1,\nonumber
\\
&~&g(r\rightarrow 0)\rightarrow 0,~g(r\rightarrow\infty)\rightarrow 1.\label{eq:g-BC}
\eea
Note that the boundary conditions of the profile function $f_{1,2}(r)$ for $r\rightarrow0$ are not necessarily equal to 0 when the corresponding winding number is 0.
Substitute the above {\it Ansatz} into \cref{tension}, we find the string tension in terms of $f_1,~f_2,~g$:
\begin{equation}
   \begin{split}
   \mu= \int d^2x \bigg[ &
      \frac{c^2 g'^2}{2r^2}+\frac{1}{2}f_1^2v_1^2\left(\frac{(j-ceg)^2}{r^2}+\frac{f_1'^2}{f_1^2}\right)
         +\frac{1}{2}f_2^2v_2^2\left(\frac{(k-ceg)^2}{r^2}+\frac{f_2'^2}{f_2^2}\right)
      \\
      &+\frac{\lambda_1}{16}(f_1^2-1)^2
         +\frac{\lambda_2}{16}(f_2^2-1)^2\bigg],
   \end{split}
\label{eq:tension1}
\end{equation}
where $'$ represents the derivative with respect to the radius $r$.

Here, we introduce an assumption about the profile functions, allowing us to derive analytical estimates for string tensions. 
More precise string profile functions are determined through numerical solutions to the equations of motion, as presented in \cref{sec:numerical_study}. 
We define critical radii, denoted as  $r_{1,c}$, $r_{2,c}$,  where the string profiles $f_1(r)$ and $f_2(r)$ respectively reach asymptotic values 
at large radii. Also, we identify $r_c$ as the radius of the magnetic flux.
A Heaviside function $\Theta(r)$ is introduced to simplify the profile functions, taking the form as 
\bea
f_1(r)&=& \Theta(r-r_{1,c})
,\quad f_2(r)=\Theta(r-r_{2,c})
,\quad
g(r)=\Theta(r-r_{c}).
\eea
Considering naturalness of the scalar mass, we take $r_c> r_{1,c},\,r_{2,c}$.
We further assume a uniform magnetic field $B_z(r)=B_0$ and evaluate the magnetic flux at the radius of $r_c$,
\be
B_0\times \pi r_c^2=\int_0^{2\pi} rd\theta Z_\theta
      =2\pi c \,  ,
\ee
yielding
\begin{equation}
   B_0=\frac{2c}{r_c^2} \, .
\end{equation}
Substituting the above expressions into \cref{eq:tension1}, we get the string tension with the winding number $(j,k)$ as
\bea
\mu&=&\frac{2\pi c^2}{r_c^2}+\pi v_1^2j^2\ln\left(\frac{r_c}{r_{1,c}}\right)+\pi v_2^2k^2\ln\left(\frac{r_c}{r_{2,c}}\right)+\pi\left(j-k\right)^2\frac{v_1^2v_2^2}{v_1^2+v_2^2}\ln\left(\frac{L}{r_{c}}\right)
\nonumber\\
&+& \frac{\pi}{16}\left(\lambda_1r_{1,c}^2v_1^4(1-\delta_{j0})+\lambda_2r_{2,c}^2v_2^4(1-\delta_{k0})\right) \, . 
\label{eq:mu_r}
\eea
We use $\int_0^L = \int_0^{r_{1(2),c}}+\int_{r_{1(2),c}}^{r_c}+\int_{r_{c}}^{L}$ when performing the radial integral.
We introduce the Kronecker delta functions, $\delta_{j0}$, $\delta_{k0}$, so that 
our result is valid for generic $j$ and $k$, including either $j,k=0$.
We can find the relation between the core sizes and the mass of the fields
by minimizing the string tension in \cref{eq:mu_r} with respect to $r_c$, $r_{1,c}$, and $r_{2,c}$ individually.
With the two Higgs masses $m_i=\sqrt{\lambda_i/2}v_i~(i=1,2)$ and the gauge boson mass $m_Z=e\sqrt{v_1^2+v_2^2}$, the radii are given as 
\bea
 r_{1,c}=\frac{2j}{m_1},\quad
 r_{2,c}=\frac{2k}{m_2},\quad
  r_c=\frac{2}{m_Z}.
\eea
Hence, the string tension can be written in terms of the mass of the massive fields
\bea
\label{eq:str-tension}
\mu&=&\frac{\pi e^2(jv_1^2+kv_2^2)^2}{2m_Z^2}+\pi v_1^2j^2\ln\left(\frac{m_1}{jm_Z}\right)+\pi v_2^2k^2\ln\left(\frac{m_2}{km_Z}\right)\nonumber\\
&+&\pi\left(j-k\right)^2\frac{e^2v_1^2v_2^2}{m_Z^2}\ln\left(\frac{m_ZL}{2}\right)+\frac{\pi}{2}\left(j^2v_1^2+k^2v_2^2\right) \, .
\eea
We define a string global charge $q_{\rm global}\equiv j-k$. 
When $q_{\rm global}=0$, the IR logarithmic divergence of the string tension $\ln(m_ZL/2)$ vanishes, implying gauge string solutions.

The full tensions of the $(1,0)$, $(0,1)$, and $(1,1)$ strings take the form 
\be
\mu_{(1,0)}=\frac{\pi v_1^4}{2(v_1^2+v_2^2)}+\pi v_1^2\ln\left(\frac{m_1}{m_Z}\right)+\pi\frac{v_1^2v_2^2}{v_1^2+v_2^2}\ln\left(\frac{m_ZL}{2}\right)+ \frac{\pi}{2}v_1^2,
\label{equ:10tension}
\ee
\be
\mu_{(0,1)}=\frac{\pi v_2^4}{2(v_1^2+v_2^2)}+\pi v_2^2\ln\left(\frac{m_2}{m_Z}\right)+\pi\frac{v_1^2v_2^2}{v_1^2+v_2^2}\ln\left(\frac{m_ZL}{2}\right)+ \frac{\pi}{2}v_2^2,
\label{equ:01tension}
\ee
\be
\mu_{(1,1)}=\frac{\pi (v_1^2+v_2^2)}{2}+\pi v_1^2\ln\left(\frac{m_1}{m_Z}\right)+\pi v_2^2\ln\left(\frac{m_2}{m_Z}\right)+ \frac{\pi}{2}(v_1^2+v_2^2).
\label{equ:11tension}
\ee
Through the above calculation, we confirm the disappearance of the IR divergent gradient energy term for a $(1,1)$ string, indicating that 
it is a gauge string.
In contrast, $(1,0)$ and $(0,1)$ strings exhibit global-like characteristics, since they have non-zero $|q_{\rm global}|=1$. 
The tension difference between gauge strings and the two global strings gives rise to the binding energy, represented by the expression
\be
\mu_{(1,0)}+\mu_{(0,1)}-\mu_{(1,1)}=\frac{\pi v_1^2v_2^2}{v_1^2+v_2^2}\left[2\ln\left(\frac{m_ZL}{2}\right)-1\right].
\label{eq:binding}
\ee
For $m_ZL\gtrsim3.3$, 
when the $(1,1)$ string string tension is less than the combined tension of $(1,0)$ and $(0,1)$ strings,
it indicates an attractive force between a $(1,0)$ string and a $(0,1)$ string. Hence, a $(1,1)$ string is more stable. 
Moreover, the hierarchical symmetry breaking scale $v_1>v_2$ implies $\mu_{(1,0)}>\mu_{(0,1)}$,
due to a larger energy stored inside the core 
of a $(1,0)$ string compared to the one of a $(0,1)$ string.

\subsection*{Generic gauge charges $q_1$ and $q_2$}
$\Phi_1$ and $\Phi_2$ can carry generic gauge charges $q_1$ and $q_2$, different from $+1$ that we assign before. 
The covariant derivative of $\Phi_1$ and $\Phi_2$ is, therefore, $D_\mu\Phi_1=(\partial_\mu-iq_1eZ_\mu)\Phi_1$ and $D_\mu\Phi_2=(\partial_\mu-iq_2eZ_\mu)\Phi_2$. 
Similarly, we deduce the string tension, taking the form
\bea
\label{eq:str-tension-q}
\mu&=&\frac{\pi e^2(jq_1v_1^2+kq_2v_2^2)^2}{2m_Z^2}+\pi v_1^2j^2\ln\left(\frac{m_1}{jm_Z}\right)+\pi v_2^2k^2\ln\left(\frac{m_2}{km_Z}\right)\nonumber\\
&+&\pi\left(jq_2-kq_1\right)^2\frac{e^2v_1^2v_2^2}{m_Z^2}\ln\left(\frac{m_ZL}{2}\right)+\frac{\pi}{2}\left(j^2v_1^2+k^2v_2^2\right),
\eea
where the string global charge $q_{\rm global}\equiv jq_2-kq_1$.

\section{Numerical Study of String Solutions \label{sec:numerical_study}}
\label{sec:3}

In this section, we present the results of our numerical investigation into the string profiles and tensions with three string solutions: $(1, 0)$ and $(0, 1)$ global strings, and $(1, 1)$ gauge string. We confirm that inside the core, the string tension of $(1, 0)$, $(0, 1)$ global strings scale with the square of the symmetry-breaking scale. Outside of the core, it exhibits logarithmic divergence.
In contrast, the tension of the $(1, 1)$ gauge string predominantly originates from its core region.

\subsubsection*{String profile}

\begin{figure}[!ht]

\begin{subfigure}{0.45\textwidth}
    \includegraphics[width=1.1\textwidth]{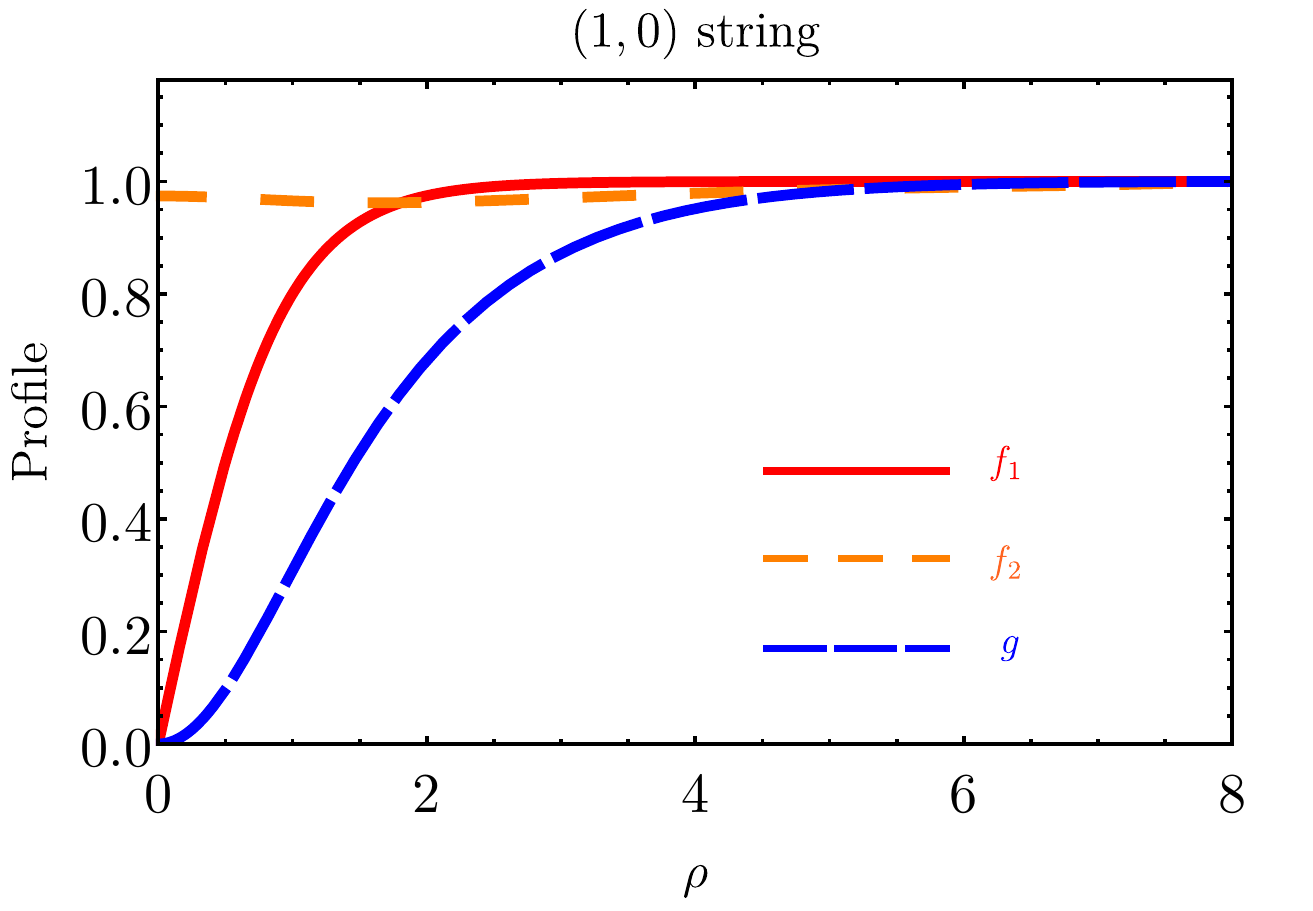}
    \centering
\end{subfigure}
\quad
\begin{subfigure}{0.45\textwidth}
    \includegraphics[width=1.1\textwidth]{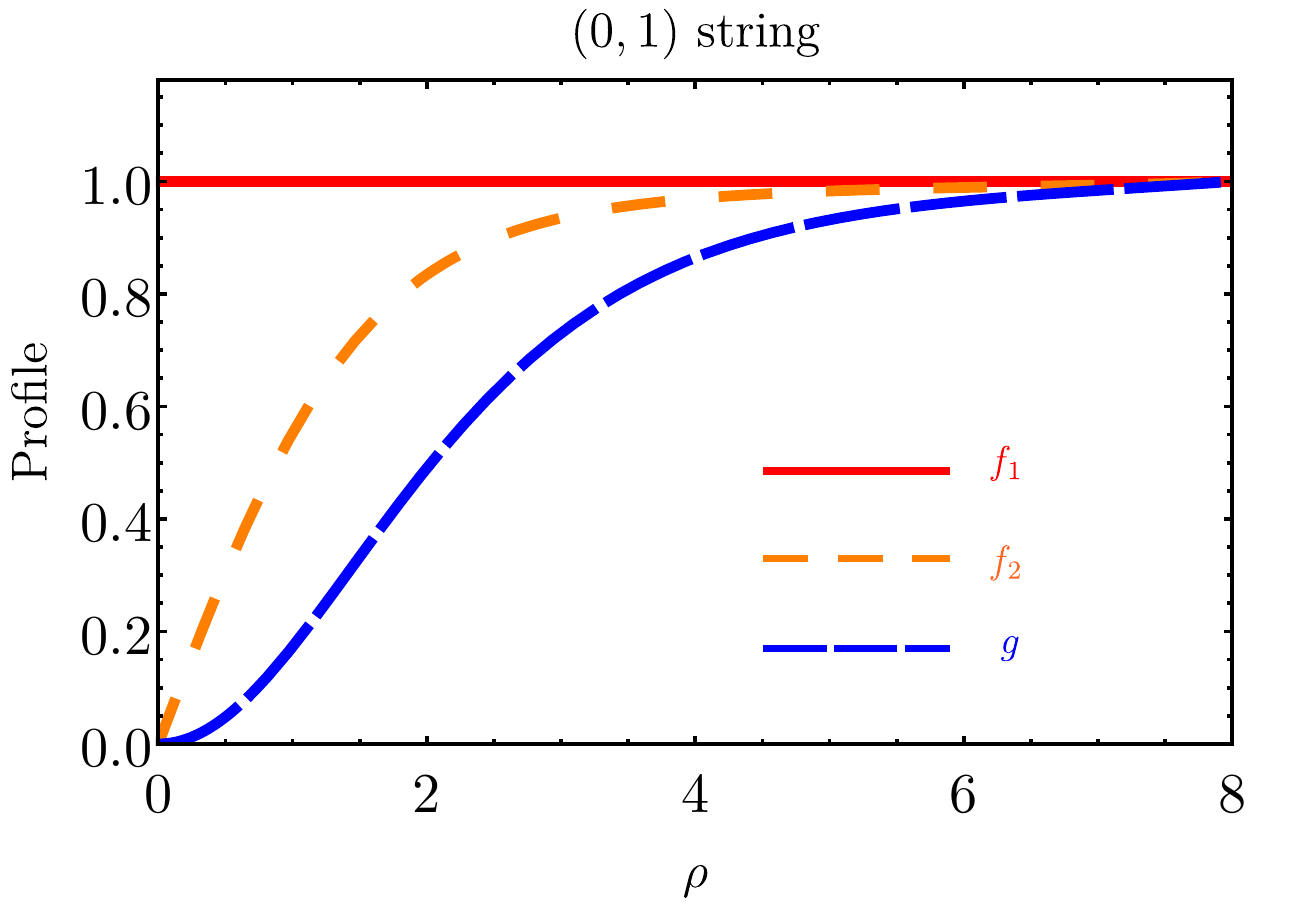}
    \centering
\end{subfigure}
\centering
\begin{subfigure}{0.45\textwidth}
    \includegraphics[width=1.1\textwidth]{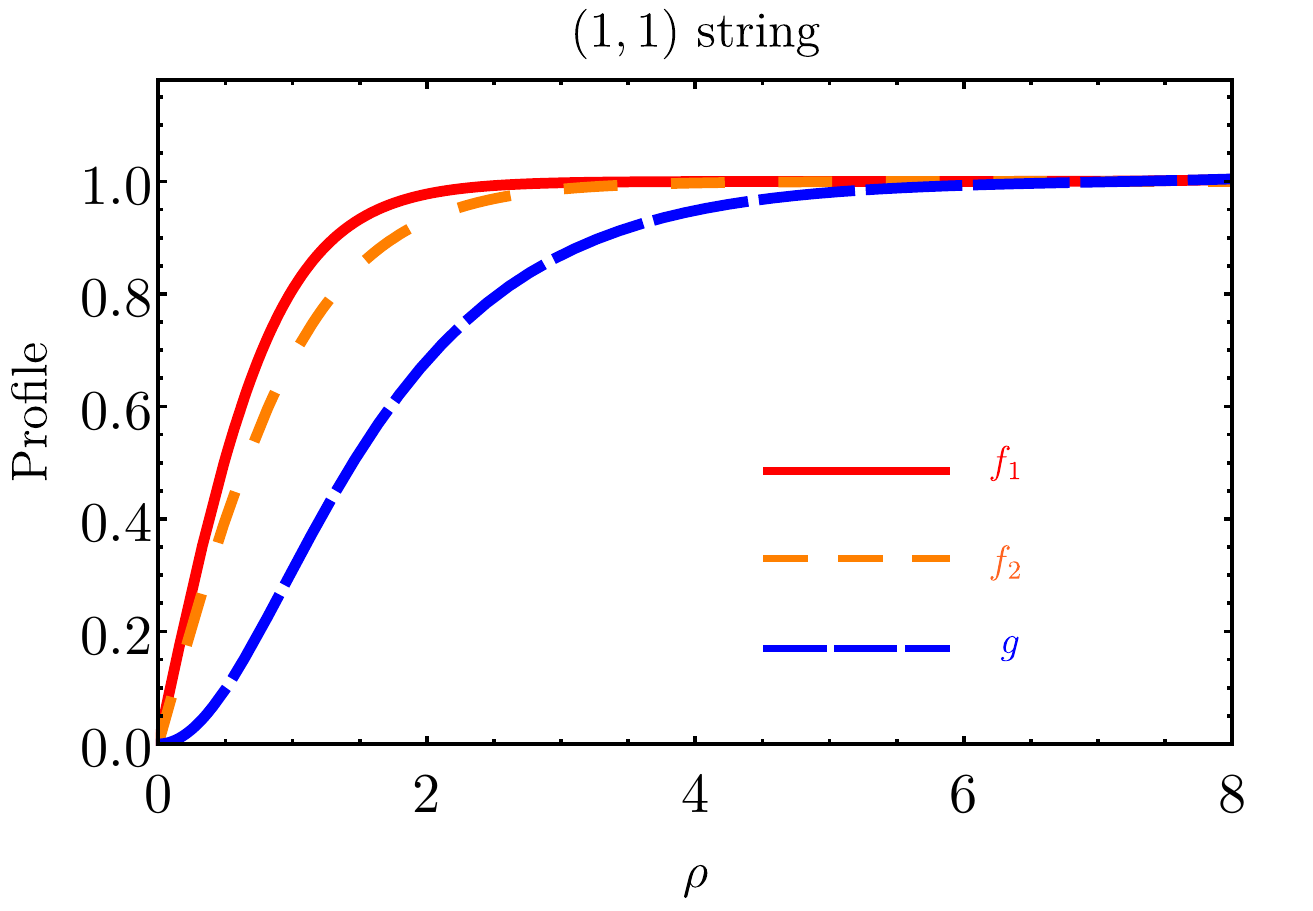}
    \centering
\end{subfigure}
    \caption{Profile functions of three string configurations as a dimensionless parameter $\rho=ev_1r$. 
}
    \label{fig: string profile}
\end{figure}

We employ the multiparameter shooting method to solve the profile functions for different string configurations with winding numbers $(j, k)$. Starting with the {\it Ansatz}, as shown in eq.\,(\ref{eq:Z-ansatz}), the equations of motion of $\Phi_1$, $\Phi_2$ and $Z^\mu$ leads to the following differential equations:
\begin{eqnarray}
    f_1''+\frac{f_1'}{r}-\frac{f_1}{r^2}(j-ceq_1g)^2+\frac{\lambda_1}{4}v_1^2(f_1^2-1)f_1=0, \nonumber \\
    f_2''+\frac{f_2'}{r}-\frac{f_2}{r^2}(k-ceq_2g)^2+\frac{\lambda_2}{4}v_2^2(f_2^2-1)f_2=0, \nonumber \\
    cg''-\frac{cg'}{r}-e^2v_1^2f_1^2(cq_1g-j)-e^2v_2^2f_2^2(cq_2g-k)=0,
        \label{equ:num}
\end{eqnarray}
where $'$ denotes the derivative with respect to $r$. These non-linear differential equations can be solved numerically by imposing boundary conditions 
at the origin, while the shooting method gives $f_1\rightarrow 1$, $f_2\rightarrow 1$, $g\rightarrow 1$ at large radius. 
The profile functions would approach zero at the origin for non-zero winding numbers due to the absence of singularity. However, for zero winding numbers, the corresponding profile functions can be non-zero values at the origin. A more detailed discussion of the profile functions for the three configurations is provided in \cref{App:B}. We consider a hierarchy in scale with $v_1 = 3v_2$. The remaining free parameters are chosen as $e=0.1, \lambda_1= 10 e^2, \lambda_2= 40 e^2,q_1=q_2=1 $. 
To simplify the analysis, we introduce the dimensionless parameter
$\rho=ev_1r$. The profile functions for $(1, 0)$, $(0, 1)$, and $(1, 1)$ strings are shown in \cref{fig: string profile}. 
More details of the numerical solutions are given in \cref{table:1} of \cref{App:B}. 
For zero winding numbers, we find that the profile functions of the corresponding scalar fields are non-zero at the origin.\footnote
{
The study in \cite{Hiramatsu:2019tua} exhibits a distinct density profile when the winding number is zero, 
where the scalar field configuration approaches zero at the origin.
Discrepancies between these two results may arise from variations in the boundary conditions 
near $r=0$ or model parameters.}

\begin{figure}[!ht]
\begin{subfigure}{0.45\textwidth}
    \includegraphics[width=1.1\textwidth]{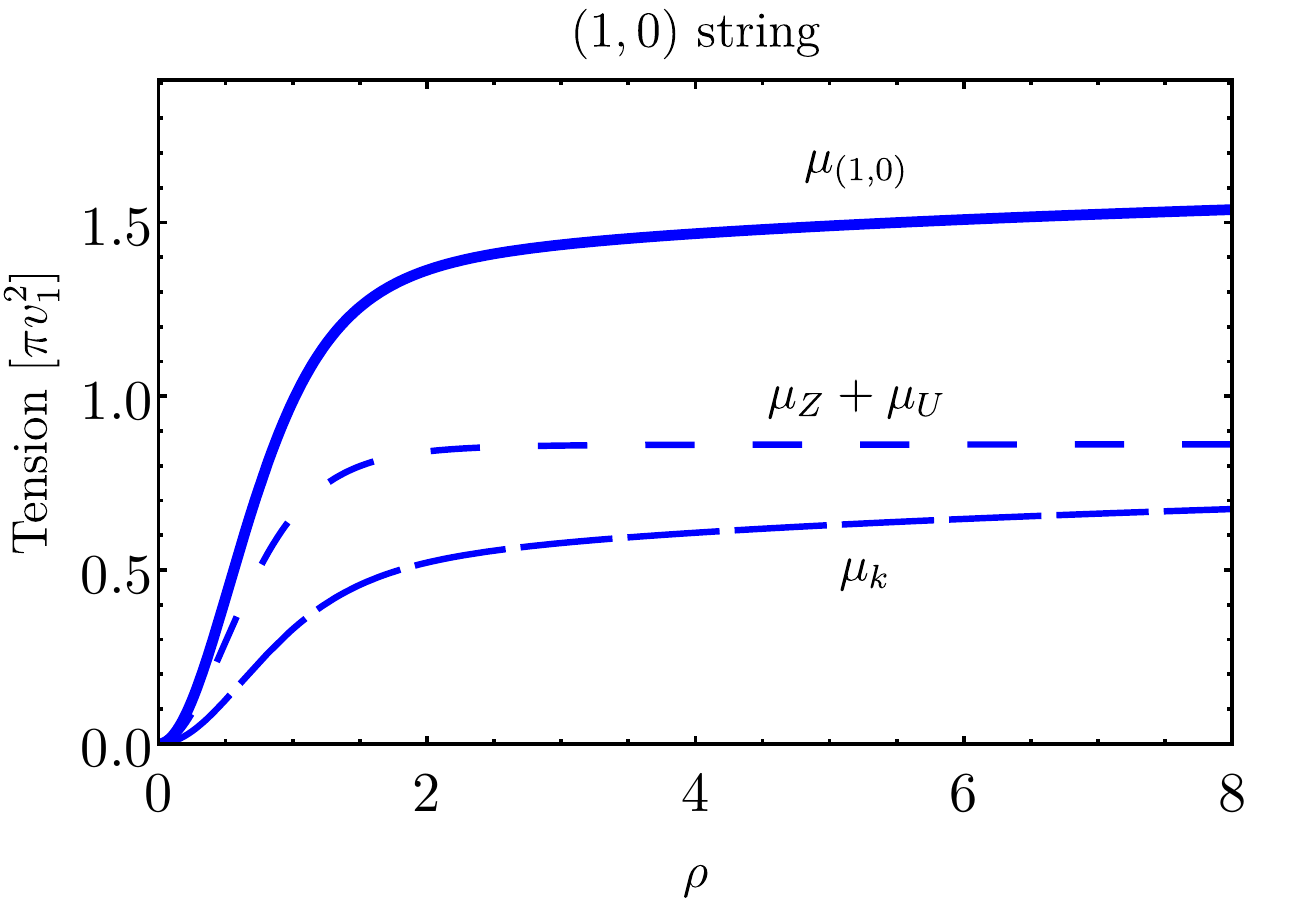}
\end{subfigure}\qquad\begin{subfigure}{0.45\textwidth}
    \includegraphics[width=1.1\textwidth]{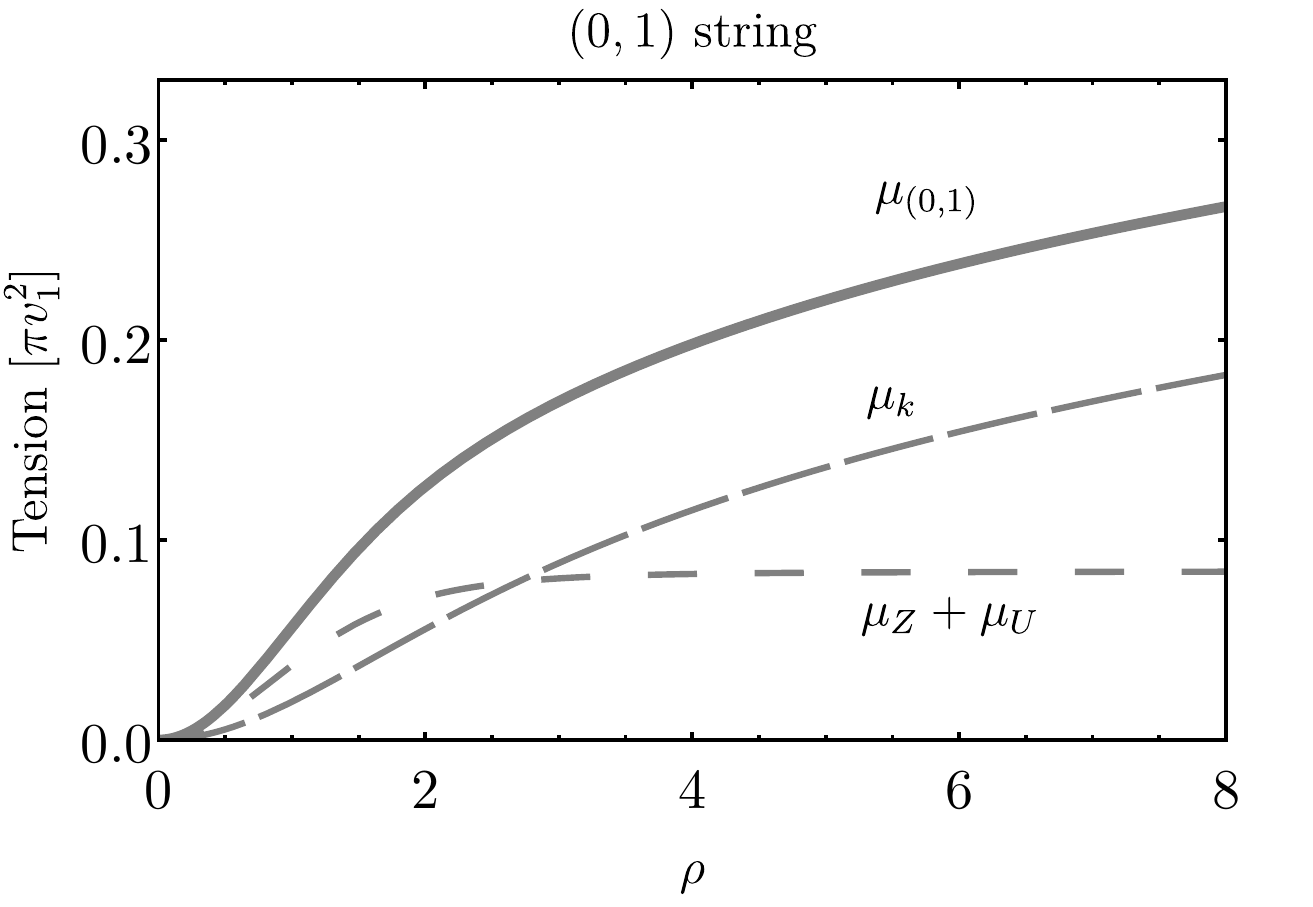}
\end{subfigure}
\begin{subfigure}{0.45\textwidth}
    \includegraphics[width=1.1\textwidth]{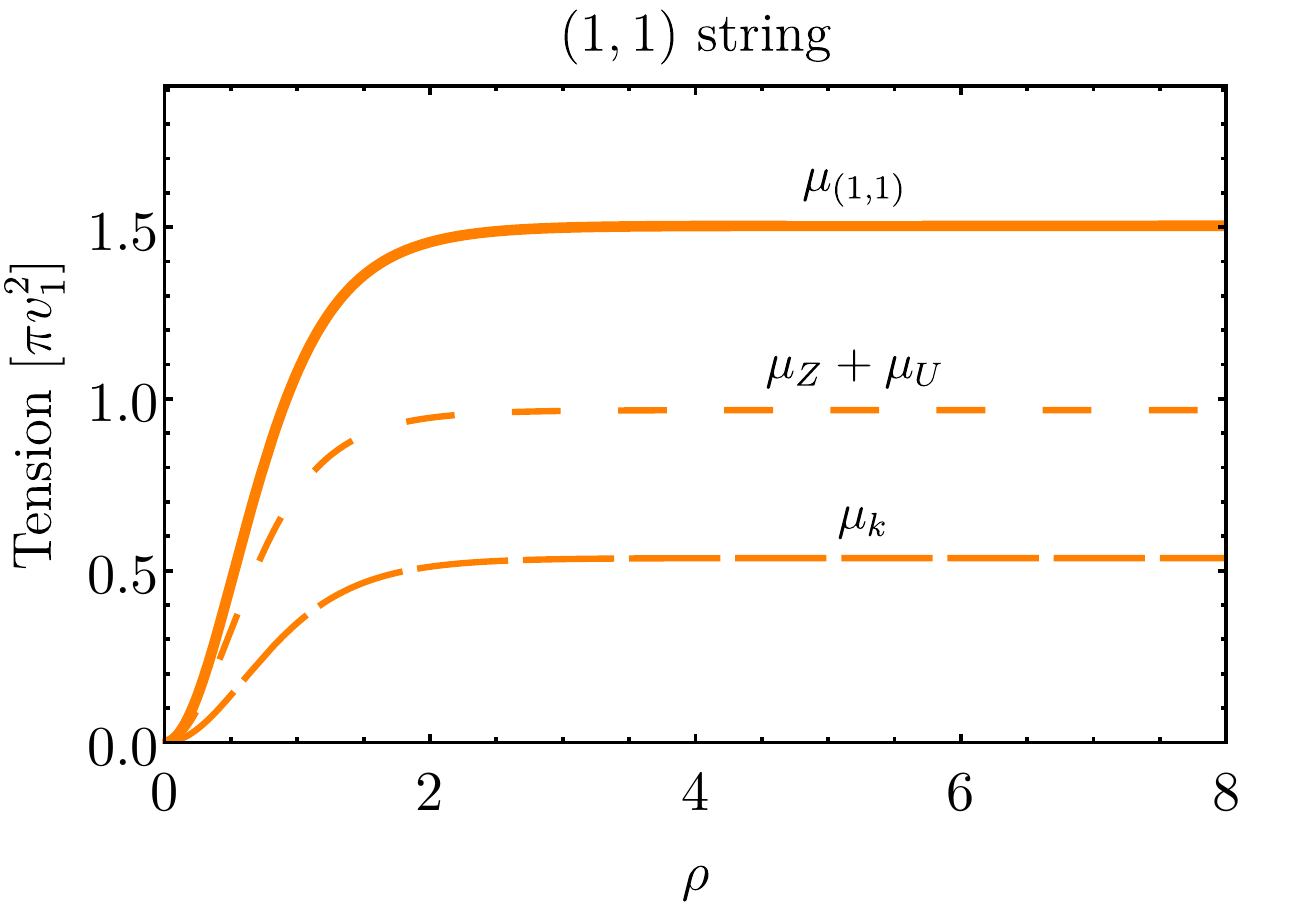}
\end{subfigure}\qquad\begin{subfigure}{0.45\textwidth}
    \includegraphics[width=1.1\textwidth]{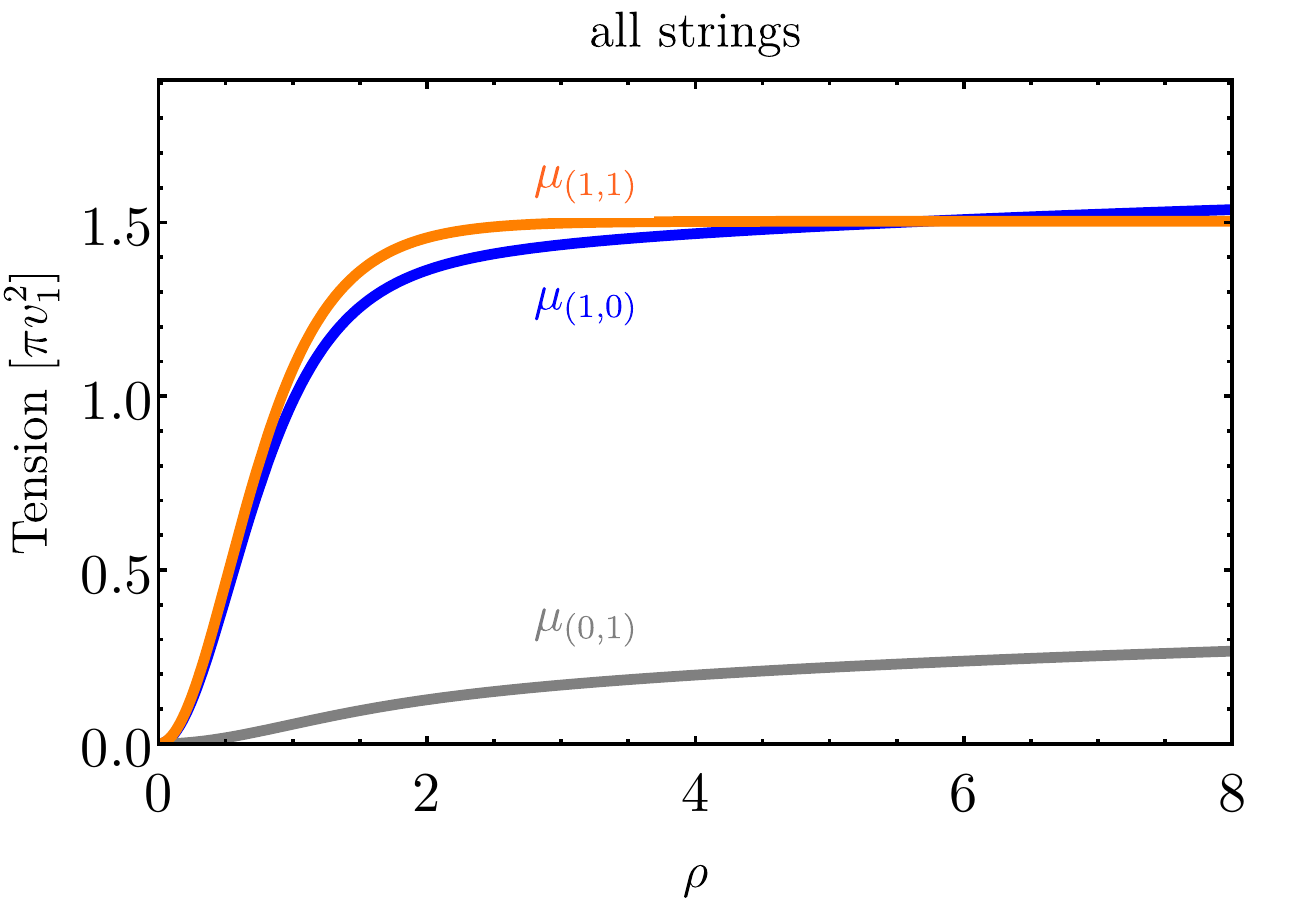}
\end{subfigure}
    \caption{String tension and its components {\it vs} $\rho$ for winding numbers $(1, 0)$, $(0, 1)$, 
and $(1, 1)$. 
The long-dashed lines represent the kinetic energy of complex scalar fields, $\mu_{k}$,
while The dashed lines represent the summation of gauge field and potential energy, $\mu_{Z} + \mu_U$.
These three components sum up to be the total tension $\mu$, as shown by solid lines.}
    \label{fig: tension profile}
\end{figure}

\subsubsection*{String tension}
We numerically compute the string tension, confirming that the $(1, 0)$ and $(0, 1)$ strings are global strings, 
while the $(1, 1)$ string is a local (gauge) string with no tension increase outside the core regime. 
Furthermore, we compare the string tension among the three configurations. 
At small scales (near the string cores), the $(1, 0)$ and $(1, 1)$ strings exhibit larger tension than the $(0, 1)$ string with 
$v_1 = 3v_2$. 
At large scales ($r>\frac{2}{m_Z}$), the tensions of $(1, 0)$ and $(0, 1)$ strings show logarithmic divergence due to the contribution from 
the kinetic energy.

We plot the string tension and its components explicitly for the three configurations in \cref{fig: tension profile}. 
The components are kinetic energy per unit length $\mu_k =  \int d^2x ( |D_i\Phi_1|^2+|D_i\Phi_2|^2 )  $, 
energy from gauge field $\mu_Z =  \int d^2x\frac{B_z^2}{2} $, and potential energy $\mu_U =  \int d^2x  V ( \Phi_1, \Phi_2) $.
Notably, 
the $(1, 0)$ and $(0, 1)$ string tension exhibit infrared logarithm divergence $\pi f_a^2 \ln( m_Z L/2)$, while the $(1, 1)$ string tension converges. 
The string separation $L$ becomes crucial in determining which string is heavier between the $(1,0)$ and the $(1,1)$.
For small separations, the $(1,0)$ string has a lower tension compared to the $(1,1)$ string, 
while for larger separations, the $(1,0)$ string becomes heavier, as shown in the last plot in \cref{fig: tension profile}.

\begin{figure}
    \includegraphics[width=0.47\textwidth]{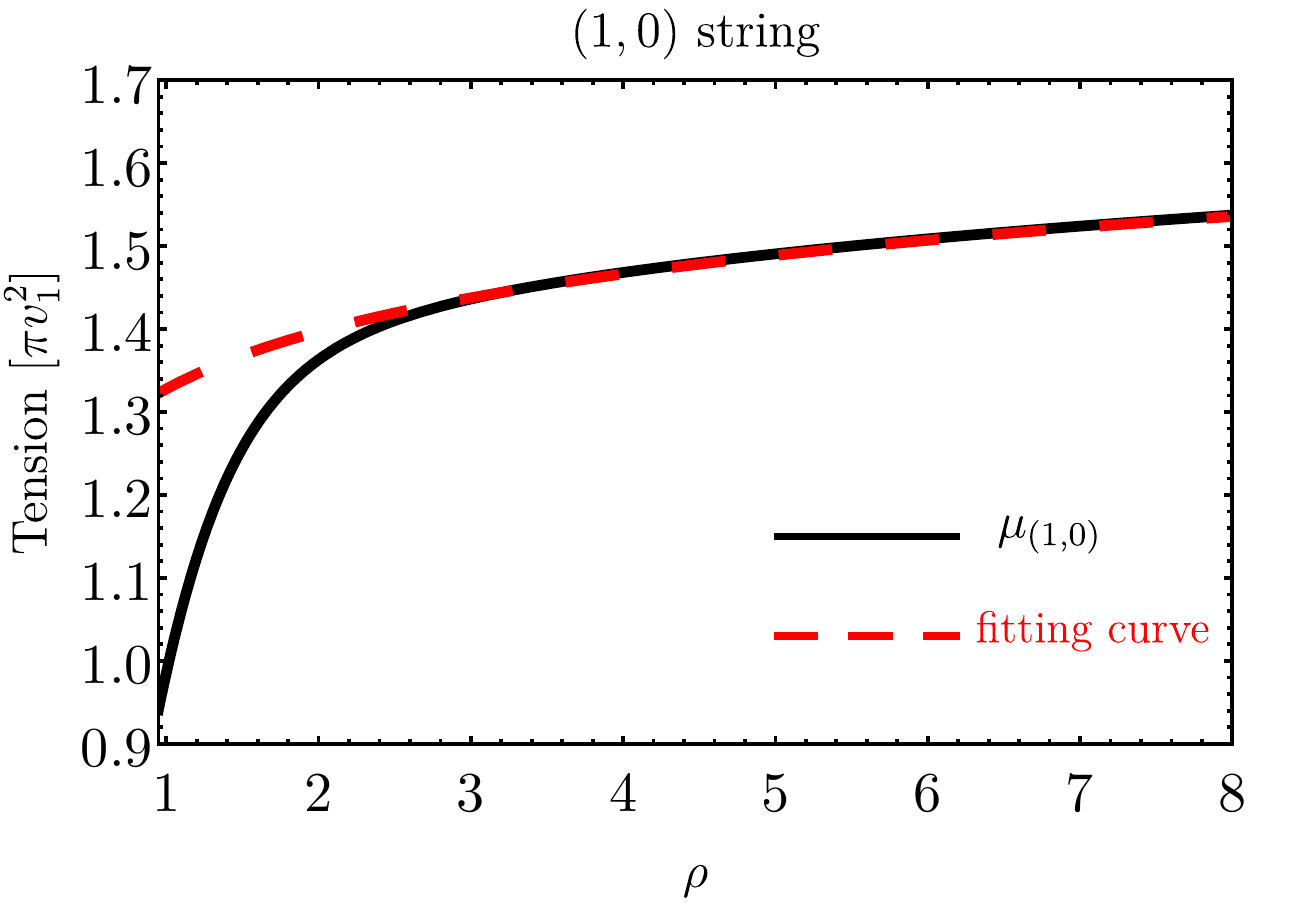}
\qquad
\includegraphics[width=0.47\textwidth]{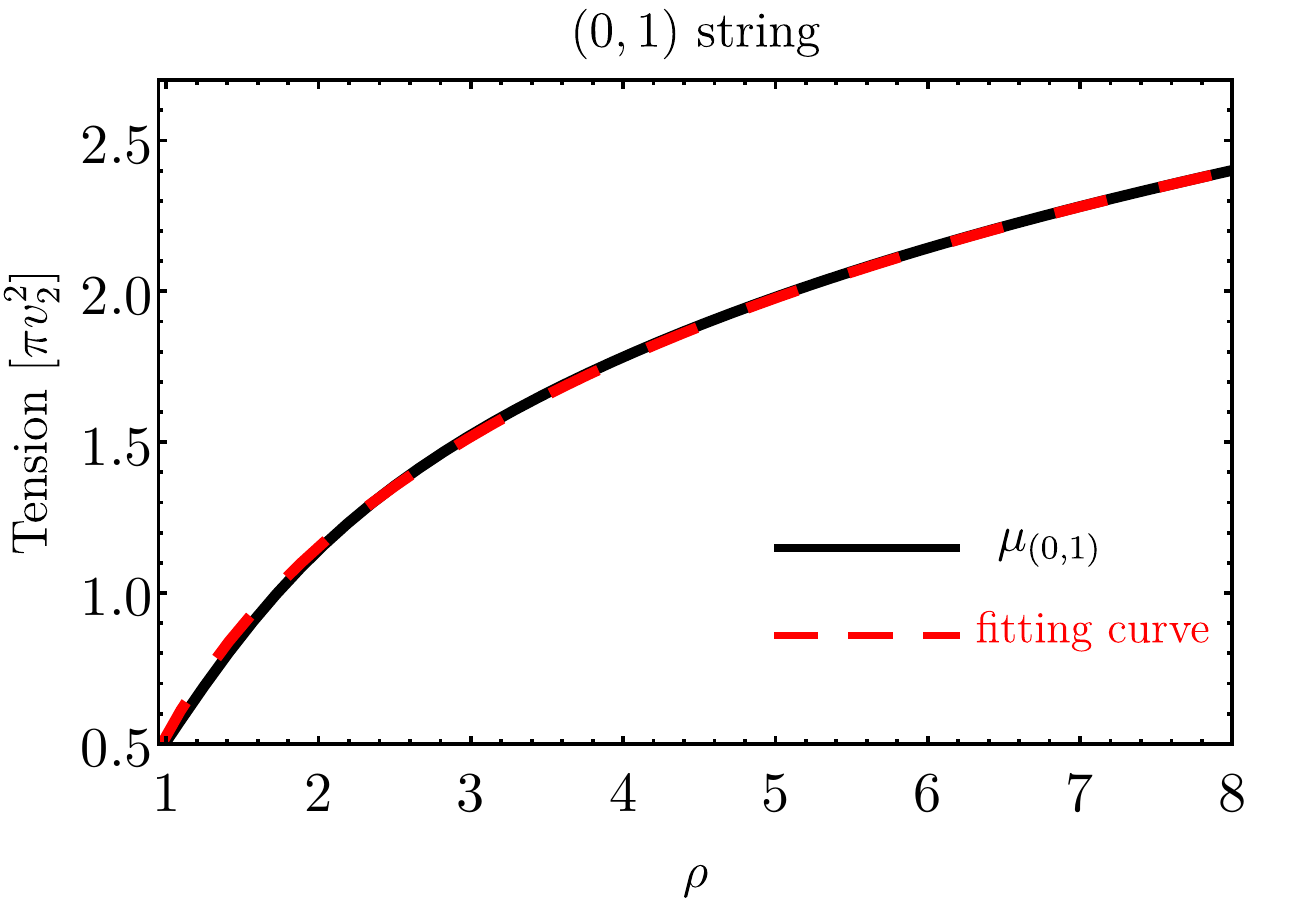}
    \caption{String tension and fitting curves in eq.\,(\ref{equ:10fit}) and eq.\,(\ref{equ:01fit}) as a dimensionless parameter $\rho$. 
}
    \label{fig: tension fit}
\end{figure}

The numerical results of string tension align with the analytic results presented in \cref{sec:str-tension}. 
The tension contribution inside the string core is of order $\pi v_1^2 \ln(\frac{m_{1}}{m_Z})$ for the $(1, 0)$ string and 
$\pi v_2^2 \ln(\frac{m_{2}}{m_Z})$ for the $(0, 1)$ string. We parameterize the $(1, 0)$ and $(0, 1)$ string tension as a combination of
string core tension plus a string tension outside the core, introducing free parameters $x_{10}$ and $x_{01}$, 
\begin{eqnarray}
    \mu_{(1,0)}(r)&=&\mu_{(1,0)}(r < \delta)+\mu_{(1,0)}(r > \delta) \nonumber\\
     &\simeq &  x_{10} \, \pi v_1^2 \ln(\frac{m_1}{m_Z})+\pi f_a^2 \ln(\frac{r m_Z}{2}), 
     \label{equ:10fit}
\end{eqnarray}
\begin{eqnarray}
    \mu_{(0,1)}(r)&=&\mu_{(0,1)}(r < \delta)+\mu_{(0,1)}(r > \delta) \nonumber\\
     &\simeq & x_{01} \, \pi v_2^2 \ln(\frac{m_2}{m_Z})+\pi f_a^2 \ln(\frac{r m_Z}{2}), 
     \label{equ:01fit}
\end{eqnarray}
where we determine $x_{10}=1.85$ and $x_{01}=3.19$ by fitting the string tension outside of the strings, as shown in \cref{fig: tension fit}.
Since $x_{10}$ and $x_{01}$ are of order $\mathcal{O}(1)$, we conclude that \cref{equ:10tension} and \cref{equ:01tension} provide the correct order 
of string tensions.

\section{Cosmological Implication\label{sec:4}}

The $\rm U(1)_Z\times U(1)_{PQ}$ model allows for the existence of two hierarchical symmetry-breaking scales, denoted as $v_1$ and $v_2$ 
($v_1 > v_2$).
The occurrence of the two $ \rm U(1)$ symmetry-breaking in the thermal history of the universe has profound implications, 
giving rise to the formation of various comic strings.
These cosmic strings consist of $(1,0)$, $(0,1)$ global strings, as well as $(1,1)$ gauge strings.
In this section, we delve into the complexities of their formation, evolution, and radiation,
which are generic and applicable to the $\rm U(1)_Z \times U(1)_{PQ}$ model and its extensions.
Moreover, we scrutinize their significance in the context of the QCD axion. 
The presence of this new string network can substantially impact axion cosmology and the calculation of axion relic abundance 
when the $\rm U(1)_Z \times U(1)_{PQ}$ symmetry is incorporated into QCD axion framework. 
Finally, we examine the radiation emitted by $(1, 1)$ gauge strings, addressing the interesting question of whether they 
predominantly emit axions or gravitational waves.

\subsection{Formation of string network\label{sec:formation-string-network}}

Two distinct phase transitions occurred during the evolution of the universe, sequentially breaking 
the $\rm U(1)_Z\times U(1)_{PQ}$.
We assume that the phase transitions are second-order, eliminating the possibility of bubble nucleation during the transition. 
The first phase transition leads to the formation of gauge string, denoted as the $\pi_1$ string, in contrast to those generated after 
the second phase transition when the symmetry is completely broken.
In the second phase transition, the $\pi_1$ strings undergo modifications
due to the scalar field $\Phi_2(x)$ configuration, and, in addition, $(0,1)$ strings form.
The interaction between these stings will establish a network of strings as the universe evolves.

During the first phase transition,
the complex scalar field $\Phi_1(x)$ acquires a non-zero VEV, $\langle\Phi_1(x)\rangle=v_1/\sqrt{2}$,
while the second scalar field $\Phi_2(x)$ remains at its potential minimum, $\Phi_2 (x) = 0 $.
As a second-order phase transition, this occurs around the temperature $T_{c,1} \sim \sqrt{\frac{6 \lambda_1}{ \lambda_1 + 3e^2}} v_1$.
The $\pi_{1}$ string with a winding number $1$ is therefore formed by the Kibble mechanism \cite{Kibble:1976sj}. 
The $\pi_{1}$ strings with 
higher winding numbers, $|j|>1$, are rarely formed. 
Since $\Phi_2(x)=0$ remains at its origin, $\pi_1$ strings can be regarded as $(1, n)$ string, with the integer $n$ determined by the 
second phase transition when $\Phi_2$ acquires a VEV.
To minimize the energy, the gauge field $Z^\mu$ compensates for the gradient of the phase of $\Phi_1(x)$, and thus the $\pi_{1}$ string
is a $\rm U(1)$ gauge string.
Following their production, these strings inevitably collide, either passing through each other or breaking and reconnecting with other strings. 
Due to these interactions, after a few Hubble times, the $\pi_{1}$ string network enters a scaling regime where a few 
long $\pi_{1}$ strings persist, and the correlation length of the long strings becomes comparable to the horizon size $H^{-1}$.

\begin{figure}
    \centering
     \includegraphics[width=0.65\textwidth]{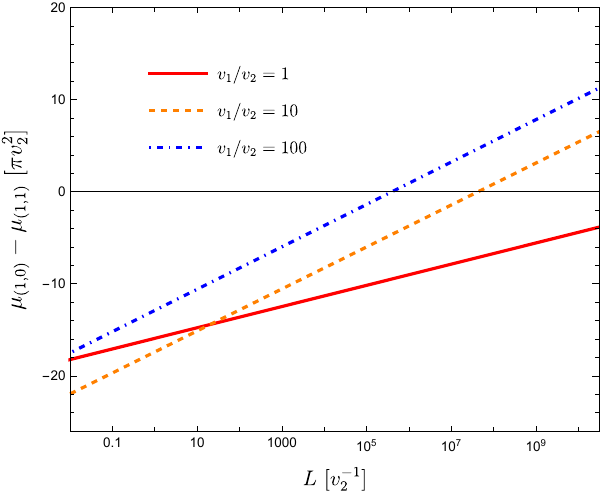}
     \caption{\label{fig:tension-ratio}The difference between $(1,0)$ and $(1,1)$ sting tension in unit of $\pi v_2^2$ as a function of $v_2L$. 
We choose 
$e=4\times 10^{-5}$, and $\lambda_1=\lambda_2=1$. $\lambda_2=1$ implies $v_2L\sim L/\xi_2$. 
}
\label{fig:Tdiff}
\end{figure}

When the temperature of the universe continuously drops below $T_{c,2} \sim \sqrt{\frac{6 \lambda_2}{ \lambda_2 + 3e^2}} v_2 $, 
another second-order phase transition occurs. 
During this transition, the complex scalar field $\Phi_2(x)$ acquires a VEV $\langle\Phi_2(x)\rangle=v_2/\sqrt{2}$. 
The correlation length of the $\Phi_2$ at the beginning of the phase transition can be estimated by the Ginzburg length \cite{Kibble:1995tx},
\be
   \xi_2\sim (\lambda_2 T_{c,2})^{-1}.
\ee
Since this correlation length is much shorter than the separation of the $\pi_{1}$ long strings, $H^{-1}$,
within the correlation length, $\Phi_1(x)$ can be considered homogeneous. Consequently, we expect the formation of 
$(0,1)$ string independently of the pre-existence of the $\pi_1$ strings.
In the case of $\pi_1$ string, within a distance $\sim\xi_2$, the presence of non-trivial winding number ($j=1$) in the $\Phi_1$ fields 
influences the $\Phi_2$ and gauge field configurations.
We assume that the field configurations adjust themselves with a slowly changing vacuum during the second-order
the phase transition to energetically favorable solutions. In this scenario, the $\Phi_2$ and gauge field reach new configurations, and 
the integer $n$ in 
the previous $(1,n)$ string is determined by minimizing the energy.
Therefore, we compare the tension $(1,0)$ to $(1,1)$ strings. The difference is estimated using \cref{equ:10tension}
and \cref{equ:11tension}, 
\begin{equation}
\mu_{(1,1)}-\mu_{(1,0)} \simeq \pi v_2^2\ln\left(\frac{m_2}{m_Z}\right)-\pi v_2^2\ln\left(\frac{m_ZL}{2}\right)=\pi v_2^2\ln\left(\frac{2m_2}{m_Z^2L}\right) \, . 
\label{eq:tension-diff}  
\end{equation}
The $(1, 0)$ strings often is lighter than $(1, 1)$ strings, particularly
considering $v_2 L \sim {\cal O} (1) $ during the second phase transition (see \cref{fig:Tdiff}).
Therefore, we expect the predominant formation of $(1,0)$ strings.   
However, it is worth noting that the phase transition may exhibit more complex dynamics, or the adiabatic argument may fail, 
potentially leading to the simultaneous production of $(1,0)$, $(0,1)$ and gauge string.

\begin{figure}
    \centering
    \includegraphics[width=0.55\textwidth]{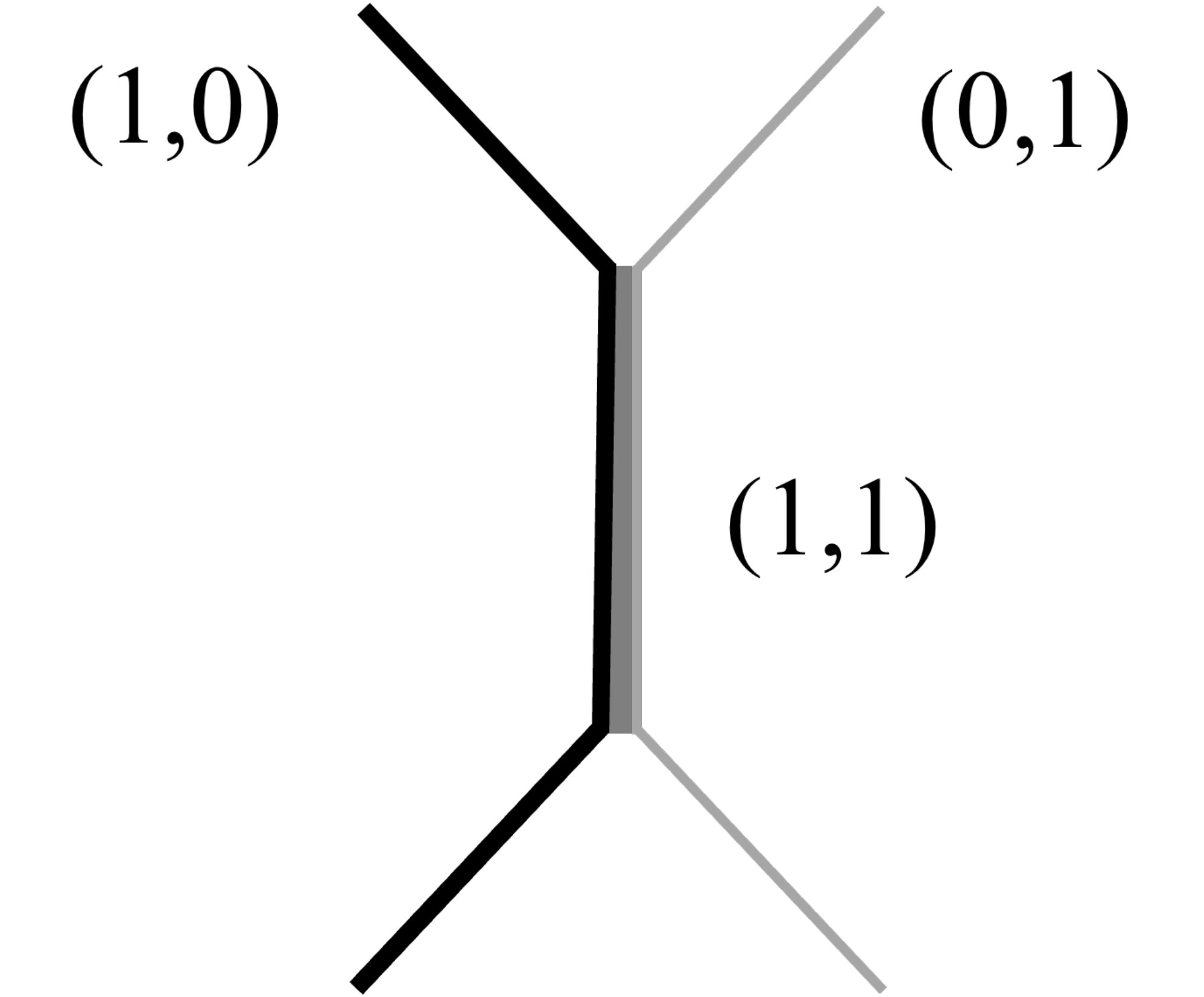}
    \caption{The sketch of a Y-junction.}
    \label{fig:y-junction}
\end{figure}

The evolution of $(1,0)$ and $(0,1)$ strings is more complicated than the evolution of conventional global string networks.
The complexity arises due to the formation of $(1,1)$ gauge string segments, known as {\it Y-junctions},   
when the two types of strings encounter and intercommute (see \cref{fig:y-junction}). 
The formation of Y-junctions can be understood as the result of attractive forces between the two strings 
or $\mu_{(1,0)}+\mu_{(0,1)}-\mu_{(1,1)} > 0 $, as shown in \cref{eq:binding}.
Although the analysis of Y-junctions requires string simulations, 
these Y-junctions cannot transform the whole $(1,0)$ and $(0,1)$ strings into a $(1,1)$ gauge string. 
First, the intersecting probability of $(1,0)$ and $(0,1)$ strings is not frequent in an expanding universe. 
Even when intersecting, they are easily unzipped by the high velocities
of the strings. Furthermore, due to the long-range interaction between a global string and its anti-string, an attractive force from  
the other side of the string loop balances the Y-junction.

These Y-junctions also occur in cosmic super-strings, with lattice simulations to analyze their evolution. 
These simulations have explored two local $\rm U(1)$ strings \cite{Urrestilla:2007yw,Bevis:2008hg,Lizarraga:2016hpd,Correia:2022spe}, global
$\rm SU(2)/Z_3$ strings \cite{Rajantie:2007hp}, and others \cite{Copeland:2005cy,Sakellariadou:2008ay,Avgoustidis:2009ke}. 
They provide insights into how Y-junctions influence the string
network. 
They consistently observe that the evolution of the string network tends towards a scaling regime. 
We anticipate similar behavior in the $\rm U(1)_Z\times U(1)_{PQ}$ model, but cosmological simulations for this model is 
encouraged to validate this conclusion.

In summary, 
we reveal that $(1,0)$ and $(0,1)$ are produced during the second phase transition
by considering the adiabatic condition. 
The $(1,0)$ and $(0,1)$ strings form Y-junctions during the evolution of the string network, 
and is expected to enter a scaling regime according to \cite{Urrestilla:2007yw,Bevis:2008hg,Lizarraga:2016hpd,Correia:2022spe,Rajantie:2007hp}.
They remain in the network with some fraction of the Y-junctions.

As a caveat, more complicated dynamics may happen, requiring dedicated cosmological simulations.
In the case that $(1,1)$ strings are generated in the second phase transition, or afterward, 
the evolution of $(1,1)$ strings is independent of $(1,0)$ and $(0,1)$ strings. 
In the string network, the $(1,1)$ string independently enters a scaling regime. Since they have the smallest tension 
compared to other strings with higher winding numbers, the $(1,1)$ strings do not bind with the $(1,0)$ or $(0,1)$ strings.
Moreover, having a generic gauge charge $q_1$ and $q_2$ may lead to gauge string formation during the 
evolution and the string network dynamics are more complicated.
The cosmological simulation conducted in \cite{Hiramatsu:2020zlp} revealed that the fraction of gauge strings could be on the same order of magnitude 
as that of global strings 
with $q_1=1,q_2=4$ and $v_1=4v_2$. The dynamics of the string network, such as the reconnection of string bundles, are highly intricate. 
The long-range force between the $(1,0)$ and $(0,1)$ strings plays an important role in forming the bound state of gauge strings when 
the charge ratio $q_2/q_1$ are large.

\subsection{QCD axion abundance}

In this section, we explore the interesting possibility of the Goldstone modes in the $\rm U(1)_Z\times U(1)_{PQ}$ being the well-motivated QCD axion. 
We consider that the symmetry breaking of $\rm U(1)_Z \times \rm U(1)_{PQ}$ occurs after inflation.
In this new scenario, we calculate the axion production. The contribution from axion vacuum realignment is the same as the predictions from 
other post-inflationary scenarios of the QCD axions.
However, the axion production resulting from the decay of the defect network, including both string and domain wall decay, can considerably
alter the axion abundance projection.
\\

\noindent
Let us present the axion production from cosmic strings and domain walls sequentially:

\subsubsection*{Cosmic strings}
Following \cite{Sikivie:2006ni}, we analyze the emission of axions from both $(1,0)$ and $(0,1)$ strings.
We exclude the contribution from $(1,1)$ strings for two primary reasons.
First, according to \cref{sec:formation-string-network}, we do not anticipate that the $(1,1)$ string number in the network is dominant. 
Second, even considering the presence of $(1,1)$ strings, the axion radiation from $(1,1)$ string loops is less than the radiation from $(1,0)$. 
This is because, during the QCD phase transition, 
the string tension of $(1,1)$ strings is smaller than that of $(1,0)$ global strings. 
This can be verified using \cref{eq:tension-diff} by setting $L\sim H^{-1}(T=\Lambda_{\rm QCD})$.
The $(1,1)$ string radiation is further discussed in  
\cref{sec:radiation-product-from-gauge-string}.
While a more complicated string network may arise, introducing uncertainties in the number density of $(1,1)$ strings, 
the findings in this section remain valid under the assumption that the scaling regime is attained.

When the string loops collapse, they convert their entire energy into axion particles. 
Consequently, the tension of the strings and the energy spectrum of the emitted axions play crucial roles in determining the axion number density. 
The tension of $(1,0)$ strings surpasses that of $(0,1)$ strings due to their heavier cores, resulting in a greater abundance of 
axion dark matter.  
The final axion density resulting from string decay also relies on the characteristics of the energy spectrum. Some attempts have been made 
to address this question through numerical simulations 
\cite{Chang:1998tb,Hagmann:2000ja, Gorghetto:2018myk, Saurabh:2020pqe,Gorghetto:2020qws,Buschmann:2021sdq}, but as far as we know, this issue has not been definitively resolved. 
Furthermore, the presence of a heavier core in the $(1,0)$ string could potentially impact the energy spectrum.
Therefore, we take an agnostic approach regarding the energy spectrum and consider two scenarios: 
one where axion emission is equally important across all energy scale (Scenario A), and 
another where emitted axions are dominated in the infrared modes (Scenario B).

As $(1,0)$ and $(0,1)$ strings enter the scaling regime, the correlation length becomes causality-limited $L\sim  t $, 
leading to the string tension of a long string taking the form 
\be
\mu_{(1,0)} (t)\simeq \pi v_1^2\ln\left(\frac{m_1}{m_Z}\right)+\pi f_a^2 \ln\left(\frac{m_Zt}{2}\right),
\label{eq:mu10}
\ee
\be
\mu_{(0,1)}(t)\simeq \pi v_2^2\ln\left(\frac{m_2}{m_Z}\right)+\pi f_a^2 \ln\left(\frac{m_Zt}{2}\right) \, .
\ee
Here, we neglect the contributions of magnetic energy and potential energy since they are subdominant and confirmed by our numerical study
as shown in eqs.~(\ref{equ:10fit}), (\ref{equ:01fit}).
Notably, the first term in \cref{eq:mu10} introduces a substantial correction to the string tension, 
owing to $v_1^2 > f_a^2 = \frac{v_1^2 v_2^2}{v_1^2+ v_2^2}$. 
In the case of $(0,1)$ strings, their tension closely resembles that of the standard QCD axion strings, as $v_2 \simeq f_a$ when $v_1 > v_2$.

Considering that the string network enters a scaling regime, the number of long strings within one Hubble patch is of the order of ${\cal O} (1)$.
The energy density of the long string within one Hubble volume at time $t$ can thus be expressed as 
\be
\rho_{\rm str, i}(t)= N_i\frac{\mu_i(t)}{t^2},
\ee
where the subscript $i$ represents the two types of strings, $(1,0)$ or $(0,1)$, and $N_i$ is the number of long strings in a Hubble patch, which is roughly $\mathcal{O}(1)$.

To consider the decay of strings into axions, we track the evolution of the density of long strings using the following equation
\be
\frac{d\rho_{\rm str,i }}{dt}+3H\rho_{\rm str, i} - H\rho_{\rm str,i} = -\frac{d\rho_{{\rm str, i}\rightarrow a}}{dt} \, ,
\label{eq:rhostr_t}
\ee
where $\frac{d\rho_{{\rm str, i}\rightarrow a}}{dt}$ is the rate at which the energy density of strings gets converted into axions
at time $t$.
The term $3H$ accounts for the dilution due to the expansion of the universe, while $-H$ arises from the stretching of the strings.  
The string decay enhances the number density of axions, resulting in the following form for the number density of axions from string decays 
\be
\frac{d n_a^{\rm str}}{dt}+3H n_a^{\rm str} =  \sum_i \frac{1}{\bar{\omega}_i(t)}\frac{d\rho_{{\rm str, i }\rightarrow a}}{dt}
       \simeq \sum_i \frac{1}{\bar{\omega}_i(t)}\frac{N_i \, \mu_i(t)}{t^3} \, . 
\label{eq:na_t}
\ee
To achieve the last approximate form, we use the string density evolution function, \cref{eq:rhostr_t}, 
and neglect the time dependence in $N_i \mu_i(t)$.
The average energy of axions, $\frac{1}{\bar{\omega}_i(t)}$, is calculated through the energy spectrum of the emitted axions, $\frac{dE_i}{d\omega}$, 
\be
\frac{1}{\bar{\omega}_i(t)}=\frac{\int \frac{1}{\omega} \frac{dE_i}{d\omega}{d\omega} }{\int \frac{dE_i}{d\omega} d\omega} \, .
\ee
The spectrum of emitted axions has some theoretical uncertainties. 
To address these uncertainties, we introduce two distinct scenarios as mentioned. 
Further, we assume that the $(1,0)$ and $(0,1)$ share the same spectrum shape to encompass these uncertainties.

In Scenario A, the spectrum peaks in the infrared region at $\omega=2\pi/t$, i.e., $\frac{dE}{d\omega}\propto\delta(\omega-2\pi t^{-1})$. 
In Scenario B, the spectrum takes the form $\frac{dE}{d\omega}\propto \frac{1}{\omega}$, with the frequency range 
$\omega\in[2\pi t^{-1},2\pi\delta^{-1}]$, 
where $\delta$ represents the UV cutoff. For a $(1,0)$ string, 
$\delta \simeq 1/m_1$; 
for a $(0,1)$ string, 
$\delta \simeq 1/m_2$.
Although we choose a delta function in the spectrum in Scenario A, the analysis should encompass
the case of $\frac{dE}{d\omega}\propto \frac{1}{\omega^q}$ with $q>1$, corresponding to an IR-dominant spectrum.
Substituting $H = \frac{1}{2 t}$, we solve \cref{eq:na_t} and find that
the number density of axions from strings can be expressed as
\be
\label{eq:str-decay}
n_a^{\rm str}(t)\simeq \frac{1}{t^{3/2}} \sum_i
      \int_{t_*}^{t}dt'\frac{1}{\bar{\omega}_i(t')}\frac{N_i\, \mu_i(t')}{t'^{3/2}} \, .
\ee
The lower limit of the time integration, $t_*$, is taken as the second phase transition time, $t(T_{c,2})$, but the
final result is not sensitive to the initial time.

In Scenario A, taking $\frac{1}{\bar{\omega}}=\frac{t}{2\pi}$ 
and neglecting the ${\cal O}(1)$ coefficients, we estimate the axion number density as follows
\begin{equation} 
      n_a^{\rm str}(t)\sim \frac{ t^{1/2}}{ t^{3/2}}   \sum_i \frac{N_i \, \mu_i(t)}{2 \pi} \, .
\label{eq:naA}
\end{equation} 
In Scenario B, where $\frac{1}{\bar{\omega}}=\frac{t}{2\pi}\left[\ln\left(\frac{t}{\delta}\right)\right]^{-1}$, the axion number density 
is estimated as 
\begin{equation}
      n_a^{\rm str}(t)\sim \frac{ t^{1/2}}{ t^{3/2}}   \sum_i \frac{N_i \, \mu_i(t) }{ 2\pi \ln\left(\frac{t}{\delta_i}\right) }  \, .
\label{eq:naB}
\end{equation}
In the two equations above, the factor of $\frac{1}{ t^{3/2}}$ accounts for the dilution due to the expansion of the universe. 
Following axion production, their number density scales as $\frac{1}{ t^{3/2}}$ in a radiation-dominant universe. 
The factor of $t^{1/2}$ indicates that 
axions produced from strings at later times dominate the axion population. Note that these formulas are valid only before the time $t_1$,
representing the horizon crossing time for the axion field, $3H(t_1)=m_a(t_1)$.
Beyond this crossing time, the IR cutoff needs to be replaced by the axion mass, resulting in the dominant axion contribution around $t_1$.
Therefore, to compute the axion density originating from the string decays, we need to evaluate the number density in 
\cref{eq:naA} or \cref{eq:naB} at the time $t_1$. As expected, the number density depends on the tension of strings and the energy spectrum.

\subsubsection*{Domain walls}
The QCD axion potential exhibits degenerate minima as the QCD axion becomes massive, leading to kinks between neighboring potential minima and 
the presence of domain walls. 
Here, we examine the QCD axion model discussed in \cref{sec:QCDaxion},  which has strings connected by a single domain wall with $N=1$.
When the axion mass turns on, each string is connected to a nearby anti-string via a domain wall. 
The domain wall solution is present in $(1,0)$ and $(0,1)$ strings. However, for $(1,1)$ strings, the PQ symmetry rotation angle 
$\alpha_{PQ} = \alpha_1 - \alpha_2 =0$, resulting in no domain wall associated with the $(1,1)$ string.

The tension of domain walls when connected to $(1,0)$ and $(0,1)$ strings is given by
\be
\sigma(t)=8m_a(t)f_{a}^2,
\ee
The axion mass $m_a(t)$ is time- or temperature-dependent. 
At high temperature ($T\gtrsim1\,{\rm GeV}$), the mass can be approximated as 
\be
m_a(T)\simeq 4\times 10^{-9}\,{\rm eV}\left(\frac{10^{12}\,{\rm GeV}}{f_a}\right)\left(\frac{{\rm GeV}}{T}\right)^4,
\ee
with the dilute instanton gas approximation \cite{Gross:1980br}, 
and at temperatures below 100 MeV
\begin{equation} 
      m_a \simeq 6 \, \mu {\rm eV} \left( \frac{10^{12} \, \rm GeV} {f_a }\right) \, .
\end{equation} 
Once the axion mass turns on at time $t_1$, the strings bound to the wall have an IR cutoff $L\sim m_a^{-1}$, comparable to the wall thickness. 
Hence the energy stored in a $(1,0)$ or $(0,1)$ string per unit length can be expressed as 
\begin{equation}
   \begin{split}
\mu_{(1,0)}(t) &\simeq \pi v_1^2\ln\left(\frac{m_1}{m_Z}\right)+\pi f_a^2\ln\left(\frac{m_Z}{2m_a(t)}\right),
      \\
\mu_{(0,1)}(t) &\simeq \pi v_2^2\ln\left(\frac{m_2}{m_Z}\right)+\pi f_a^2\ln\left(\frac{m_Z}{2m_a(t)}\right).
   \end{split}
\label{eq:stringTt2}
\end{equation}

After the axion mass turns on at $t_1$, the domain walls become bound to the strings. 
The evolution of the domain walls goes through several stages.
Initially, the dynamics of the wall-string system are governed by the strings until time $t_2$, and the scaling solution of long strings remains;
subsequently, starting from time $t_2$, the domain walls begin pulling the strings, and the dynamics of the system become dominated by the walls.
Finally, at time $t_3$, the domain walls and strings collapse, emitting axion particles.

The transition time $t_2$
is determined when the energy stored in the strings is comparable to the energy in the walls,
\be
   \frac{E_\sigma(t_2)}{E_\mu(t_2)} \simeq \frac{8m_a(t_2) f_a^2  t_2^2}{\mu(t_2) t_2}=1.
\label{eq:Et2}
\ee
The $(1,0)$ and $(0,1)$ strings have different transition times due to their distinct string tension $\mu(t)$, as shown in \cref{eq:stringTt2}.
For walls connected to the light $(0,1)$ strings, the situation aligns with the standard scenario \cite{Chang:1998tb}, 
which approximates $t_2\sim t_1$.
However, when the walls are connected to the heavier $(1,0)$ strings, where $\mu_{(1,0)}>\mu_{(0,1)}$, 
it takes longer to achieve the energy balance between the wall and the string, giving us $t_2>t_1$.

Beyond $t_2$, the energy stored in the wall surpasses that stored in the strings bound to the wall due to their different scaling with time.
The walls pull the strings and accelerate them. 
The strings eventually unzip the wall into several smaller walls until the wall's size is comparable to $1/m_a$.
Finally, the walls collapse, emitting axions at $t_3$.

The energy density of the walls at time $t_1$ was estimated to be approximately $0.7$ per horizon volume \cite{Chang:1998tb}. 
This energy density scales as $\sigma(t)/t$ when the dynamics are governed by the strings.
After $t_2$, the domain wall area does not change significantly within one Hubble volume, so the average energy density  
is scaled by the volume or the inverse cubic power of the cosmological scale factor 
$t^{-3/2}$, 
\be
   \rho_{\text {DW}}(t)\sim 0.7 \frac{\sigma(t)}{t_2}\left(\frac{t_2}{t}\right)^{3/2}, \quad t_2 < t< t_3 \, .
\label{eq:rhoDW}
\ee
After time $t_3$, 
the axions produced during the collapse of the wall-string system are boosted.
We define an average Lorentz factor $\gamma$ as
the ratio of the energy density to the axion mass at $t_3$, denoted as $\gamma \equiv \frac{\langle\omega\rangle}{m_a(t_3)}$.
The number density of the produced axions is scaled by the volume
\be
\label{eq:DW-decay}
   n_a^{\rm DW}(t)\sim \frac{\rho_{\rm DW}(t_3)}{\langle\omega\rangle}\left(\frac{t_3}{t}\right)^{3/2}
   \sim \frac{6}{\gamma}\frac{f^2_{a}}{t_2}\left(\frac{t_2}{t}\right)^{3/2} \, .
\ee
In the second equality, substituting \cref{eq:rhoDW}, we find that the dependence on time $t_3$ drops out of the estimate of $n_a^{\rm DW}(t)$. 
According to simulation results in Ref.~\cite{Chang:1998tb}, $\gamma\sim 60$, though there is significant uncertainty on this value.

We consider the $(1,0)$ string and its wall-string system here since it has large tension and contributes significantly to the axion density, 
but in the numerical analysis, we include contributions from both global strings.
We compare the axion number density produced by the $(1,0)$ string bound to the wall using \cref{eq:DW-decay} with the one produced by the $(1,0)$ string decay in Scenario A and B.
The domain wall contribution to axion number density does not depend on $t_3$, as shown in \cref{eq:DW-decay},   
allowing us to compare the string decay contribution with the string-wall collapse contribution at $t_2$ directly.
In Scenario A, 
we compare $n_a^{\rm str}$ with $n_a^{\rm DW}$ at $t_2$, and set the number of $(1,0)$ long strings is $N_{(1,0)}\sim 1$,
\begin{equation}
\begin{split}
n_a^{\rm str}(t_2)&=n_a^{\rm str}(t_1)\left(\frac{t_1}{t_2}\right)^{3/2}
\sim \frac{\mu}{m_a(t_1)t_2^2}\left(\frac{t_2}{t_1}\right)^{1/2}
      \\
      &> \frac{\mu}{m_a(t_1)t_2^2}\sim \gamma n_a^{\rm DW}(t_2).
\end{split}
\end{equation}
In the last line, we use \cref{eq:Et2} to replace $\mu(t_2)$.
Considering $t2> t_1$ and $\gamma \gtrsim 1$, 
the axion production from the wall decay is sub-dominant in Scenario A.
In Scenario B, the axion spectrum is harder, leading to a smaller number density of axion from strings.
Numerically, 
we find that the domain wall contribution remains sub-dominant in most regions of the model's parameter space even when $\gamma=1$.

There is one more complication stemming from Y-junctions. The bound $(1,1)$ strings do not connect with the domain walls. 
The collapse of global strings with the Y-junctions is expected to close the $(1,1)$ strings and leave $(1,1)$ string loops in the universe. 
However, the $(1,1)$ strings, even when they eventually emit axions, represent a sub-dominant contribution to the axion dark matter abundance, 
and thus, they are neglected in our analysis.

\subsubsection*{Results}
The total axion energy density at present, $t=t_0$, is given by the sum of the three contributions: misalignment, string decays, and 
domain wall decays,
\be
\rho_{a,0}=\rho^{\rm vac}_a(t_0)+m_{a}n_a^{\rm str}(t_0)+m_{a}n_a^{\rm DW}(t_0) \, .
\ee
$n_a^{\rm str}$ is the number density of axions from both $(1,0)$ and $(0,1)$ string decay.
There is an uncertainty of the number of long strings $N_i$ in a Hubble patch. Here, we just set $N_i=1$ as an estimation.
$n_a^{\rm DW}$ is the number density of axions from domain wall decay bounded by $(1,0)$ and $(0,1)$ strings, 
and $\rho^{\rm vac}_a=\Omega_a^{\rm vac}\rho_{c,0}$
is the energy density of axions produced by the misalignment mechanism,
with $\rho_{c,0}$ the current critical density of the universe, for $f_a<2\times10^{15}~$GeV, 
\be
\Omega_a^{\rm vac} h^2\sim 2\times 10^4\left(\frac{f_a}{10^{16}\text{ GeV}}\right)^{7/6}\langle\theta_{a}^2\rangle,
\ee
where $\langle\theta_{a}^2\rangle=\pi^2/3$ is the average value of the square of the misalignment angle.

\begin{figure}
    \centering
    \includegraphics[width=0.75\textwidth]{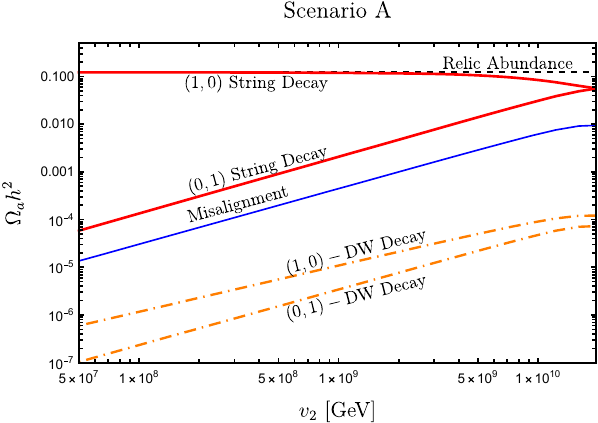} \\  
    \vspace{0.5cm}
    \includegraphics[width=0.75\textwidth]{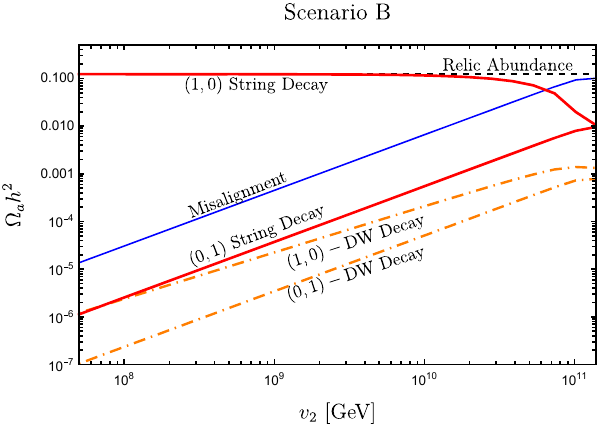} 
    
    \caption{\label{fig:Omega_a}The three contributions to the dark matter relic abundance as a function of $v_2$. 
The red solid lines show the string decay contribution from $(1,0)$ and $(0,1)$ strings. The blue solid line shows the misalignment mechanism contribution. The orange dot-dashed lines show the domain wall decay contribution from the walls bound by $(1,0)$ and $(0,1)$ strings. The total axion abundance is shown by a horizontal black dashed line. We set the gauge coupling $e=4\times 10^{-5}$, $\lambda_1=\lambda_2=1$, the Lorentz factor $\gamma=60$ for domain wall decay.}
    
\end{figure}

If one considers axion to explain the 100\% of the relic abundance of dark matter observed today, this will require
\be
\Omega_ah^2\equiv\frac{\rho_{a,0}}{\rho_{c,0}}h^2=\Omega_{\rm DM}h^2\sim 0.12.
\ee
\Cref{fig:Omega_a} summarizes the three contributions to axion energy density as a function of $v_2$ for both Scenario A and B. It confirms that
the decay from the $(1,0)$ string with the heavy core is the dominant contribution to dark matter relic abundance for most regions of $v_2$.  

\begin{figure}
    \centering
     \includegraphics[width=0.75\textwidth]{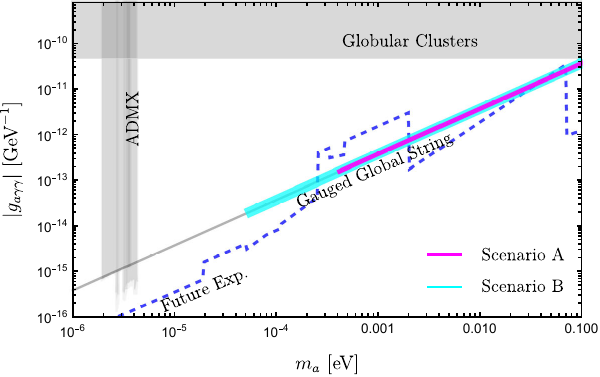}\\
      \vspace{0.5cm}
     \includegraphics[width=0.72\textwidth]{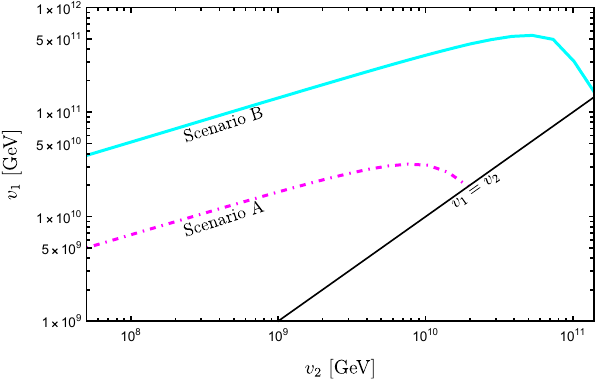}
     \caption{\label{fig:axion_bound}The bound on the QCD axion-photon coupling as a function of the axion mass (Top). The existing constraints are shown by gray-shaded regions from ADMX experiments \cite{ADMX:2009iij,ADMX:2018gho,ADMX:2019uok,ADMX:2021nhd} and globular clusters \cite{Dolan:2022kul}. The combined projection using future haloscope experiments is shown by a blue dashed line \cite{Stern:2016bbw,Alesini:2020vny,DeMiguel:2023nmz,Lawson:2019brd,Beurthey:2020yuq,McAllister:2017lkb,Aja:2022csb,brass,BREAD:2021tpx}. The gray solid line denotes the power-law dependence of the axion-photon coupling on QCD axion mass in the KSVZ model. Considering the QCD axion to explain 100\% dark matter relic abundance, the magenta and cyan bands show the mass region opened up by the gauge global string model in Scenarios A and B. 
     The bottom panel shows the value of $v_1$ and $v_2$ to reproduce the dark matter relic abundance. The lines are truncated at $v_1=v_2$ since we assume $v_1\gtrsim v_2$.
     }
\end{figure}

In the KSVZ model, the axion-photon coupling is linked to the axion mass through the relation 
$g_{a\gamma\gamma}=2.0\times10^{-16}C_{a\gamma}(m_a/\mu{\rm eV})\,{\rm GeV}^{-1}$, with $C_{a\gamma}=-1.92$. 
Given the results obtained above for making QCD axions to explain the 100\% of the dark matter relic abundance, 
we show the axion-photon coupling versus the axion mass predicted by our model in \cref{fig:axion_bound} (Top), 
see the magenta (Scenario A) and cyan (Scenario B) bands. 
The current ADMX experiments \cite{ADMX:2009iij,ADMX:2018gho,ADMX:2019uok,ADMX:2021nhd} exclude the QCD axion mass range $2-4\,\mu {\rm eV}$, 
while the globular cluster observations \cite{Dolan:2022kul} put an upper bound on the axion mass around 0.1\,eV.

We do not analyze mass below $\mu {\rm eV}$, as they are associated with the pre-inflation scenario, where strings do not form. 
According to a recent axion string simulation \cite{Buschmann:2021sdq}, the axion mass is found to be within the range of 
$m_a\in(40,180)\,\mu {\rm eV}$. 
In contrast, our proposed mechanism opens a large mass window for QCD axions as 100\% cold dark matter. This offers motivation for the upcoming haloscope 
experiments \cite{Stern:2016bbw,Alesini:2020vny,DeMiguel:2023nmz,Lawson:2019brd,Beurthey:2020yuq,McAllister:2017lkb,Aja:2022csb,brass,BREAD:2021tpx} 
to probe axions across a broad mass spectrum.

\subsection{$(1,1)$ gauge string radiation: gravitons or axions?\label{sec:radiation-product-from-gauge-string}}

For macroscopic gauge strings, the production of massive particles is significantly suppressed, 
while the emission of massless gravitons becomes the dominant channel for these strings to lose energy. 
However, the $(1,1)$ gauge string not only couple to massless gravitons but also to axions.
This section aims to address whether the $(1,1)$ gauge strings predominantly decay into axions or gravitons.

To comprehend the interaction between strings and axions, we start with the Lagrangian density of the two scalar fields, 
as shown in \cref{eq:Lag}, and investigate the axion interactions with classical configurations of $Z^\mu(x)$, $\Phi_1(x)$ and $\Phi_2(x)$.
Subsequently, we perform a gauge transformation, $Z^\mu \to Z^\mu  + \frac{1}{e} \partial_\mu \alpha_Z$.
This transformation eliminates the quadratic term of $Z^\mu \partial_\mu a$ and $Z^\mu \partial_\mu \pi_z$.
The Lagrangian density thus takes the form
\begin{equation}
   {\cal L}  = - \frac{1}{4} Z_{\mu \nu } Z^{\mu \nu } + \frac{1}{2} e^2 ( \phi_1^2 + \phi_2^2 ) Z_\mu^2
      -  g(\phi_1, \phi_2) \, e \, Z^\mu \partial_\mu a + \frac{1}{2} \, f(\phi_1, \phi_2) \, (\partial_\mu a)^2   \, .
\label{eq:Lv2}
\end{equation}
In this equation, we introduce the functions $g(\phi_1, \phi_2)$ and $f(\phi_1, \phi_2)$ to simplify the notation,
\begin{equation}
   g(\phi_1, \phi_2) =   f_a \frac{ \phi_1^2 }{v_1^2}   -  f_a \frac{ \phi_2^2 }{v_2^2 }   \, ,  
\end{equation}
\begin{equation}
   f( \phi_1 , \phi_2) = \frac{  g^2( \phi_1 , \phi_2)  + \frac{1}{f_a^2} \phi_1^2 \phi_2^2 }{ \phi_1^2 + \phi_2^2 }    \, .
\end{equation}
When considering the vacuum expectation values of $\phi_1$ and $\phi_2$, $g(v_1, v_2) = 0 $ and $f( v_1, v_2) =1 $.
This implies that there is no interaction between axion and classical fields outside string cores.
\Cref{eq:Lv2} already reveals that $(1,1)$ strings interact with axions through the term $ g(\phi_1, \phi_2) \, e \, Z^\mu \partial_\mu a$, 
owing to the non-trivial field configurations of $Z^\mu(x)$, $\phi_1(x)$, $\phi_2(x)$. One example of these field configurations of $(1,1)$ string 
is shown in \cref{fig: tension profile}, where the classical fields gradually change in the core size of $\sim 1/ m_Z$.

To calculate the radiation power of $(1,1)$ strings to axions, we can employ the Kalb-Ramond field $B^{\mu\nu}$ \cite{Kalb:1974yc,
Vilenkin:1986ku,Davis:1988rw}.
A duality relationship exists 
between real massless field, axion, and the two-form antisymmetry gauge field, $B^{\mu \nu}$, given by 
\begin{equation} 
   \partial_\mu a  = \frac{1}{2} \epsilon_{\mu \nu \alpha \beta } \partial^\nu B^{\alpha \beta } \, 
\end{equation} 
This relation is satisfied beyond the string core.
However, in the presence of $\phi_1(x)$, $\phi_2(x)$ and $Z^\mu (x)$, this duality relationship is modified.
Based on the field equation of $a(x)$ derived from \cref{eq:Lv2}, 
\begin{equation}
    \partial^\mu ( f( \phi_1 , \phi_2) \, \partial_\mu a  - g( \phi_1 , \phi_2)  \, e Z_\mu ) = 0 \, ,
\end{equation}
the relation between $a$ and $B^{\mu \nu}$ should be replaced with
\begin{equation}
      f( \phi_1 , \phi_2) \, \partial_\mu a  - g( \phi_1 , \phi_2)  \, e Z_\mu 
      = \frac{1}{2} \epsilon_{\mu \nu \alpha \beta } \partial^\nu B^{\alpha \beta } \, 
\end{equation}
Substituting axion with $B^{\alpha\beta}$ or its gauge-invariant field strength $H_{\mu \nu \lambda }$ in the Lagrangian density, 
we obtain the following expression,  
\begin{equation}
   \begin{split}
   {\cal L} &= 
    - \frac{1}{4} Z_{\mu \nu } Z^{\mu \nu } + \frac{1}{2} e^2 ( \phi_1^2 + \phi_2^2 ) Z_\mu^2
      -\frac{1}{2} e^2 \frac{ g(\phi_1, \phi_2)^2 } { f(\phi_1, \phi_2) } Z_\mu^2 
      \\
      &+ \frac{1}{12} \frac{1}{f(\phi_1, \phi_2) } H_{\mu \alpha\beta} H^{\mu \alpha \beta}
      + \frac{1}{2} \epsilon_{\mu \nu \alpha \beta} B^{\alpha \beta} \partial^\mu \left(    \frac{ g(\phi_1, \phi_2) } { f(\phi_1, \phi_2) } e Z^\nu 
      \right)
     \, . 
   \end{split}
\label{eq:L2form}
\end{equation}
where the field strength of $B^{\mu \nu}$ is defined as, 
\begin{equation}
     H_{\mu \alpha \beta }     = \partial_\mu B_{\alpha \beta } + \partial_\alpha B_{\beta \mu} + \partial_\beta B_{\mu \alpha} \, .
\end{equation}
The last term in each line of \cref{eq:L2form} is included to ensure that 
the field equations for $Z^\mu$ and $a$ remain consistent with the original Lagrangian. 
The field equation for $H^{\nu \alpha \beta}$ is given by 
\begin{equation}
   \partial^\nu  \frac{ H_{\nu \alpha \beta} } { f(\phi_1, \phi_2) } 
      +  \epsilon_{\mu \nu \alpha \beta} \, \partial^\mu \left(   \frac{ g(\phi_1, \phi_2) } { f(\phi_1, \phi_2) } e Z^\nu  
         \right) 
      =  \epsilon_{\mu \nu \alpha \beta} \partial^\mu \partial^\nu a    
\end{equation}
The right-hand side of this equation is zero in the vacuum or the background of $(1,1)$ string but is connected to the winding number 
in the global strings, leading to nontrivial coupling between the global strings and axions. Although the gauge $(1,1)$ string lacks coupling 
between the winding number flux and $B^{\mu \nu }$, the 
coupling to $Z^\mu$ term as shown in the last term in \cref{eq:L2form} leads to the radiation of axion from the string. 
The action for the interaction takes the form
\begin{equation}
   S_j = 
      \int d^4 x \, \frac{1}{2} B^{\mu \nu } j_{\mu\nu} 
      =   
      \int d^4 x \, 
      \frac{1}{2} 
      \epsilon_{\mu \nu \alpha \beta} 
      B^{\alpha\beta} 
   \partial^\mu \left(   \frac{ g(\phi_1, \phi_2) } { f(\phi_1, \phi_2) } e Z^\nu \right)
     \, .
\end{equation}
Analogous to the derivation of the gravitational wave radiation power \cite{Weinberg:1972kfs}, 
the radiation power per solid angle at a frequency $\omega$ in a direction ${\bf k}$ 
is proportional to the amplitude square of the Fourier transformation of $j^{\mu \nu}$, 
\begin{equation}
   \frac{ {\rm d} P_a}{ {\rm d} \Omega} = 
        \frac{  \omega^2} { 32 \pi^2 } 
       j^{\mu \nu * }  (\omega, {\bf k})  
          j_{\mu \nu  }  ( \omega, {\bf k} )  
\end{equation}
where $|{\bf k}| = \omega $ for massless axions, and 
\begin{equation} 
   j_{\mu \nu  }  ( \omega, {\bf k}) = \int d^4 x  \, e^{- i k \cdot  x} j_{\mu \nu  }  ( x ) \, .
\end{equation} 
We then estimate the radiation power as 
\begin{equation}
   \frac{ {\rm d} P_a}{ {\rm d} \Omega} \sim e^2 f_a^2 \, .
\end{equation}
This estimation considers that the classical field $Z^\mu$ varies on the order of $m_Z$ within a core of size on the order of $1/m_Z$.
The radiation power of global strings is approximately $\sim f_a^2$, and the radiation power of gauge strings to graviton is $\sim G f_a^4$, with $G$ the Newtonian gravitational constant. 
In contrast, the radiation to axions from $(1,1)$ strings is smaller than that from global strings due to the gauge coupling suppression, $e^2$. 
Nevertheless, the
radiation of axions is dominant over that of gravitons, which is suppressed by the Planck mass unless an exceedingly smaller gauge coupling is 
considered. For the case when the gauge coupling is highly suppressed, we estimate that the gravitational wave spectrum plateau produced by $(1,1)$ 
string loop oscillations is $\Omega_{\rm GW}(f)\sim O(10^{-4}) \sqrt{{G \mu}/{\Gamma}}$, with the constant efficiency of 
gravitational wave emission $\Gamma\sim 50$.

\section{Conclusion}
\label{sec:conclusion}

In conclusion, we investigate the string solutions and cosmological implications within the ${\rm U}(1)_{\rm Z}\times{\rm U(1)_{PQ}}$ model. 
This model has two hierarchical symmetry-breaking scales given by two complex scalar fields.
By assuming that both $\rm U(1)$ symmetries spontaneously broke after the inflation, we find that the ${\rm U}(1)_{\rm Z}\times{\rm U(1)_{PQ}}$ model 
can give a rich feature of string networks in cosmology, produce more axions as cold dark matter when embedding the model into the QCD axion
framework, and, further, have a new decay product from the gauge string.

We identify three distinct types of string solutions: $(1,0)$ global string with a heavy core, $(0,1)$ global string with the tension close to 
QCD axion strings, and $(1,1)$ gauge string as a bound state of the former two global strings. 
Numerical studies confirm the existence of these string solutions, providing a solid foundation for our cosmological considerations.

In the early universe, the formation of string networks during the second phase transition predominantly yield $(1,0)$ strings instead of $(1,1)$ 
gauge strings, assuming an adiabatic process to generate these strings. Due to the attractive interaction between $(1,0)$ and $(0,1)$ strings, 
it allows the global strings to form a bound state of $(1,1)$ gauge strings during the evolution of the string network, 
such that Y-junctions of three types of strings are expected to be present in this string network.

Furthermore, we introduce a KSVZ-like model and make the Nambu-Goldstone mode become the QCD axion. 
We find that 
the decay of $(1,0)$ strings with heavy cores dominates axion radiation, 
providing a potential explanation for the observed dark matter relic abundance for the masses exceeding $10^5 \, {\rm eV}$.
Additionally, the $(1,1)$ gauge string in this model has a coupling with axions due to the spatial dependent profile of the gauge 
field $Z^\mu$ within the string core. 
This presents a novel decay channel of gauge strings into axions.

While our work initiates the exploration of this hybrid cosmic strings network and its QCD axion implication, 
it raises several interesting questions requiring string simulations.
It includes the evolution of Y-junctions, the existence of scaling solutions, and the energy spectrum of QCD axion radiated by $(1,0)$ strings.
Also, the preference of $(1,1)$ gauge strings to radiate axions or gravitons awaits confirmation through numerical studies.
Simulations of the $\rm U(1)_Z \times U(1)_{PQ}$ model with a large gauge charge ratio 
\cite{Hiramatsu:2020zlp}, as well as 
studies that replicate the axion cosmic strings with heavy core global strings \cite{Klaer:2017qhr}
have been performed, but many of the questions raised above remain unanswered.

The rich phenomenology of this model offers testable predictions.  
Upon improving the detection sensitivity of future haloscope experiments, the parameter space in $g_{a\gamma\gamma}-m_a$ predicted by this model 
in the QCD axion framework can be probed. Additionally, if the gauge coupling is small enough, the gravitational radiation is predominantly produced 
by the $(1,1)$ gauge string. The existence of Y-junctions in the string network can modify the loop distribution function compared to conventional
cosmic string scenarios \cite{Brandenberger:2008ni}. Therefore, the gauge strings could yield a distinctive gravitational wave signal from this model.

\section*{Acknowledgments}
We thank 
Pierre Sikivie, Robert Brandenberger, Jeff Dror, Keisuke Harigaya, Sungwoo Hong, Rachel Houtz, Junwu Huang, Subir Sarkar, Sergey Sibiryakov, and Tanmay Vachaspati for valuable discussions. This work was supported in part by the U.S. Department of Energy under grant DE-SC0022148 at the University of Florida.

\appendix
\section{Asymptotic value of the gauge field configuration\label{App:A}}

The asymptotic value of $Z_\theta(r)=\frac{c}{r}$ at large $r$ can be obtained by requiring $\frac{\partial\mu}{\partial c}=0$. At large $r$, we write down the string tension with only $c$-dependent terms, i.e., the kinetic terms,
\bea
\mu&=&\int d^2x\frac{1}{2}\left[f_1^2v_1^2\left(\frac{(j-q_1 e c)^2}{r^2}+\frac{f_1'^2}{f_1^2}\right)+f_2^2v_2^2\left(\frac{(k-q_2 e c)^2}{r^2}+\frac{f_2'^2}{f_2^2}\right)\right]\nonumber\\
&=&\pi\int dr r\left[v_1^2\left(\frac{(j-q_1 e c)^2}{r^2}\right)+v_2^2\left(\frac{(k-q_2 e c)^2}{r^2}\right)\right]\nonumber\\
&=&\pi\left[v_1^2(j-q_1 e c)^2+v_2^2(k-q_2 e c)^2\right]\int \frac{dr}{r} \, . 
\eea
In the second equality, we use $f_{1,2} \to 1$ and $f_{1,2}^\prime \to 0 $. 
By minimizing the string tension with respect to $c$, we can find the value of $c$:
\bea
\frac{\partial\mu}{\partial c}&=&2\pi\left[(v_1^2q_1^2+v_2^2q_2^2)ec-(jv_1^2q_1+kv_2^2q_2)\right]\int \frac{dr}{r} =0\\
\Rightarrow &c&=\frac{1}{e}\frac{jv_1^2q_1+kv_2^2q_2}{v_1^2q_1^2+v_2^2q_2^2}.
\eea

\section{Numerical solutions to the string profile functions\label{App:B}}

We take large-$\rho$ behaviors $f_1\rightarrow 1$, $f_2\rightarrow 1$, $g\rightarrow 1$ are satisfied at $\rho=8$ in our numerical study, and \cref{fig: string profile}) confirms that the profile functions capture the asymptotic values in all configurations. The small $\rho$ behaviors can be found by taking $\rho\rightarrow 0$ in the equation of motions \cref{equ:num}. For $(1, 0)$ string, $f_1\rightarrow0$ and $g\rightarrow0$ must be satisfied at the origin because of no singularity. However, $f_2$ doesn't need to be so. Explicitly, the boundary conditions for a $(1, 0)$ string takes 
the form 
\begin{equation}
\begin{split}
    \lim_{\rho \to \epsilon}f_1(\rho) &= c_1+\beta_1 \rho + O(\rho^2) 
      \\
    \lim_{\rho \to \epsilon}f_2(\rho) &= c_2  + \beta_2 \rho^2 + O(\rho^3)
      \\
    \lim_{\rho \to \epsilon}g(\rho) &= m \rho^2+O(\rho^3) \, ,
\end{split}
\end{equation}
so that we take six boundary conditions from the values at $\rho = \epsilon $ and their derivatives. 
Here $c_1=0$, $\beta_2= \frac{\lambda_2}{16}(c_2^2-1)c_2$, and we take $\epsilon=10^{-3}$. The left three parameters $\beta_1$, $c_2$, and $m$ are determined by the shooting method at $\rho=8$. Similarly, the $(0,1)$ string shares the same large $\rho$ asymptotic behaviors as the $(1,0)$ string. The boundary conditions for $(0, 1)$ string at $\rho\rightarrow 0$ are
\begin{equation}
\begin{split}
    \lim_{\rho \to \epsilon}f_1(\rho) &= c_1  + \beta_1 \rho^2 + O(\rho^3)
         \\
    \lim_{\rho \to \epsilon}f_2(\rho) &= c_2+\beta_2 \rho + O(\rho^2)
      \\
    \lim_{\rho \to \epsilon}g(\rho) &= m \rho^2+O(\rho^3) 
\end{split}
\end{equation}
where $c_2=0$ and $\beta_1=\frac{\lambda_1}{16}(c_1^2-1)c_1$.
The profile functions of the $(1, 1)$ string must satisfy the smoothness at the origin:
\begin{equation}
\begin{split}
    \lim_{\rho \to \epsilon}f_1(\rho) &= c_1+\beta_1 \rho + O(\rho^2) \\
    \lim_{\rho \to \epsilon}f_2(\rho) &= c_2+\beta_2 \rho + O(\rho^2)\\
    \lim_{\rho \to \epsilon}g(\rho) &= m \rho^2+O(\rho^3)  \, ,
\end{split}
\end{equation}
where $c_1=c_2=0$ due to no singularity at the origin. Under the parameter space mentioned in \cref{sec:3}, we find the numerical result 
in \cref{table:1}. 

\begin{table}[h] 
    \centering 
    \caption{Numerical result of string profile functions.}
\begin{tabular}{|c|c c c c c|}
\hline
configurations& $\beta_1$ & $\beta_2$  & $c_1$ & $c_2$ & $m$\\
\hline
$(1, 0)$ string & 1.1025 & -0.0012 & 0 &0.9741 &0.4248\\

$(0, 1)$ string & $-1.24\times 10^{-6}$ & 0.6302& 0.9999& 0 &0.1992\\

$(1, 1)$ string & $1.12143$ & 0.847589& 0 & 0 &0.428086\\
\hline
\end{tabular}
    \label{table:1}
\end{table}

\bibliographystyle{JHEP}
\bibliography{ref}
\end{document}